\documentclass[aps,nobibnotes,prb,superscriptaddress,twocolumn,floatfix,showpacs,amsmath,amssymb, amsfonts,mathtools]{revtex4-1}

\usepackage{graphicx}
\usepackage[pdftex]{hyperref}
\usepackage{subcaption}
\usepackage{floatrow}
\usepackage{latexsym}
\usepackage{psfrag}
\usepackage[makeroom]{cancel}
\usepackage{dcolumn}
\usepackage{blindtext}
\usepackage{epstopdf}
\floatstyle{plaintop}
\restylefloat{table}

\usepackage{amsmath,latexsym,psfrag}
\usepackage{array}
\usepackage{dcolumn}
\usepackage{bm}
\usepackage{soul}  
\usepackage{gensymb}

\begin{document}
	\title{Low Born effective Charges, High Covalency and Strong Optical Activity in $X_3^{2+}$Bi$^{3-}$N$^{3-}$ ($X$=Ca,Sr,Ba) $inverse$-perovskites}
	\author{Jasmine Wakini}
	\affiliation{ Theoretical Condensed Matter Group (TCMG), Department of Natural Science, 
	The Catholic University of Eastern Africa,
	62157 - 00200, Nairobi, Kenya.}
	\author{Carolyne Songa}
\affiliation{ Theoretical Condensed Matter Group (TCMG), Department of Natural Science, 
	The Catholic University of Eastern Africa,
	62157 - 00200, Nairobi, Kenya.}
\author{ Stephen Chege}
\affiliation{ Materials Modeling Group (MMG), 
	School of Physics and Earth Science,
	The Technical University of Kenya,
	52428-00200, Nairobi, Kenya.}
		
	\author{Felix O. Saouma}
	\affiliation{
		Computational and Theoretical Physics Group (CTheP), Department of Physical sciences,
		Kaimosi Friends University College,
		385-50309, Kaimosi, Kenya.}
	\author{Elica Wabululu}
\affiliation{ Theoretical Condensed Matter Group (TCMG), Department of Natural Science, 
	The Catholic University of Eastern Africa,
	62157 - 00200, Nairobi, Kenya.}
		\author{P.W.O Nyawere}
	\affiliation{
		Physical and Biological Sciences Department, P.O BOX Private Bag 20157, Kabarak, Kenya}
		\author{Victor Odari}
\affiliation{ Masinde Muliro University of Science and Technology,
	190-50100, Kakamega, Kenya.} 
\author{ James Sifuna}
	\email[For correspondence:~]{jsifun@cuea.edu}
\affiliation{ Theoretical Condensed Matter Group (TCMG), Department of Natural Science, 
	The Catholic University of Eastern Africa,
	62157 - 00200, Nairobi, Kenya.}
\affiliation{ Materials Modeling Group (MMG), 
	School of Physics and Earth Science,
	The Technical University of Kenya,
	52428-00200, Nairobi, Kenya.}
		\author{George S. Manyali}
	\affiliation{
		Computational and Theoretical Physics Group (CTheP), Department of Physical sciences,
		Kaimosi Friends University College,
		385-50309, Kaimosi, Kenya.}
	\date{\today}
	
	\begin{abstract}
We compute for the first time a complete charge analysis (Bader and Born effective) on $X_3^{2+}$Bi$^{3-}$N$^{3-}$ ($X$=Ca,Sr,Ba). The crystals show a great electron sharing with little possibility of ferroelectricity. $Inverse$ perovskites have been a center of attraction in the recent years and not much is known on the systems under this study. This research addressed some key missing components and decomponents like the hardness and optical spectrum in $X_3^{2+}$Bi$^{3-}$N$^{3-}$ ($X$=Ca,Sr,Ba). The computed lattices slightly deviated from the parent perovskites indicating a future interfacing under a proper substrate. We also found out that all the crystals under this study were semiconducting  with direct band gaps but plastic in nature due to strong covalency. The optical spectrum revealed very strong activity in these crystals in the ultraviolet regime. The information herein will definitely guide the experimentalist in fabrication of these materials for novel functionalities.

	\end{abstract}
	
	\maketitle
	\section{INTRODUCTION}
	\label{sec:introduction}
 Research on \texorpdfstring{$inverse$}-perovskites~\cite{Anti-perovskites1,PhysRev,PhysRevB.99.205126,CHEPKOECH2020e00484,Japan,Kariyado_PRB,KAUR2020101741,Nuss:dk5032} is in a momentous stage. These materials are very interesting and can be compared to the rapture of semiconductor physics many years ago~\cite{when_oxides_meet}. This is a genuine comparison since our lives are already oscillating on devices developed from $inverse$ perovskites~\cite{future_antiperovskites}. Among the subtle physics recently discovered in some $inverse$ perovskites include; magnetism~\cite{Magnetism_ant-perovskites,Magnetsm_anti1,Magnetsm_anti2}, excellent electrical and thermal conductivities~\cite{electrical_thermal_conductivities}, superconductivity~\cite{superconductivity_1}, adjustable thermal expansion~\cite{thermal_expansion}, energy conversion mechanism~\cite{energy_conversion}, ferromagnetic shape memory effects~\cite{TAKENAKA2013S291}, High order topology~\cite{High-order-topology-Anti-perovskites} and energy storage capabilities~\cite{energy_storage}. 

\begin{figure}[H]
	\centering
	\includegraphics [width=\columnwidth]{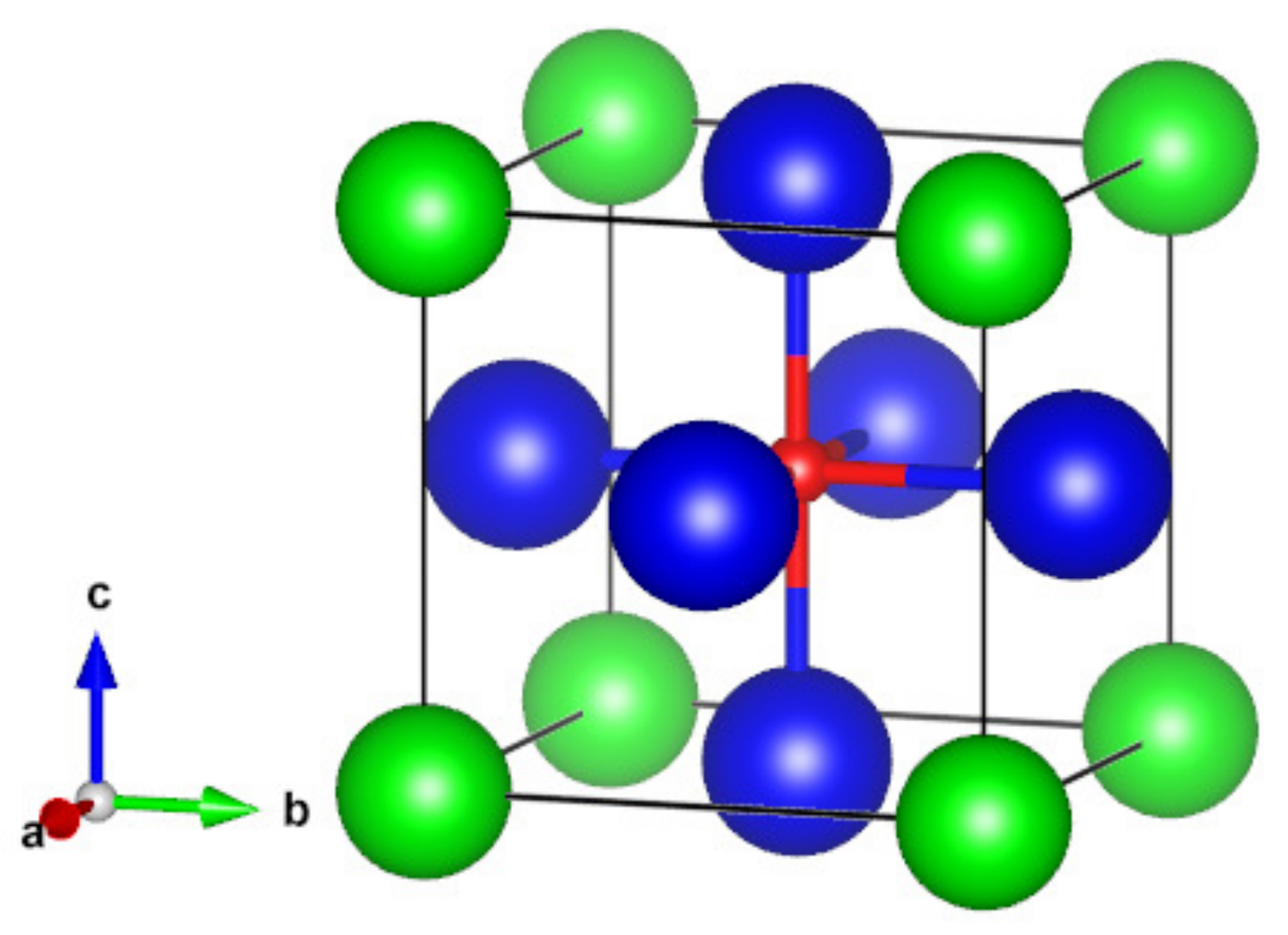}
	\caption {(Color online) The Crystal structure of the $inverse$-perovskite $X_3^{2+}$Bi$^{3-}$N$^{3-}$ ($X$=Ca,Sr,Ba) under this study. The green  balls represent Bi$^{3-}$, blue balls represent the $X_3^{2+}$ cations while the red ball represent N$^{3-}$ anion. }
	\label{fig:Crystal}
\end{figure}

 In a clear manifestation, inverted or $inverse$-perovskite nitrides are typically inorganic crystals with a perovskite looking structure with a transpose in cations and anions and thus deviating from the analogous perovskite family as explained in Ref.~\onlinecite{Evans2020_antiperovskite}. Taking a look at Fig.~\ref{fig:Crystal}, we note that the selected $inverse$ perovskites under investigation, crystallize in $Pm\overline{3}m$ space group of No. 221 in which the N atom occupies the center of the cube surrounded by an octahedrally coordinated six $X$ atoms while the Bi atoms being found at the vertices of the cells. 
 Other reports show that these materials can exist in orthorhombic~\cite{Ullah_2016} and hexagonal~\cite{hexagonal_anti-perov} phases. Ref.~\onlinecite{semiconducting_inverse-perovs} hints at a possibility of  Sr$_3$BiN and Ba$_3$BiN having a semiconducting nature, We anticipate that Ca$_3$BiN will also show semiconducting traits since the key contributing orbitals to conductivity are the same all through. Some other extensive work on the selected crystals has been on the optical spectrum. Ref.~\onlinecite{Ullah_2016} depicts in a nut shell as to what is expected as far the spectrum is concerned and we wondered if we can obtain the same spectrum but by employing HSE06. The authors~\cite{Ullah_2016} report of a very strong optical activity in all the selected compounds under this study and we sort to really understand the regime in which this novel behaviour occurs so as to tailor its application.  There is also scanty information as far as the mechanical stability of these crystals is concerned. Ref.~\onlinecite{hexagonal_anti-perov} and Ref.\onlinecite{Ullah_2016} indicate that these promising materials are plastic in the cubic phase.

We can see that there are attempts to fully characterize these materials among the scientific community. However, much still needs to be done. To this date, we still have glaring gaps in literature pointing towards these novel $inverse$ perovskites. It is most likely that these literature gaps do hinder the technological exploration in the $X_3^{2+}$Bi$^{3-}$N$^{3-}$ ($X$=Ca,Sr,Ba) $inverse$-perovskites. One trait that stands out to be typically missing as far as these materials are concerned is the concept of the Bader and  Born effective charges. Charge analysis has not been reported yet and this greatly limits ferroelectric exploration in such crystals. Charge analysis helps to ascertain the bonds in a crystal and predict a possibility of future crystal engineering. On the same note, we don't have a very clear road map as far as the optical spectrum is defined. We understand the DFT bandgap problem that typically affects the optical characterization. These has led to non-converged optical spectrum in literature. We hope to seal this loophole by use of the implemented HSE06 scheme in the {\sc Siesta} method~\cite{SIESTA}. It has been pointed out that our selected materials have a weak bond but unfortunately, it is not yet clear as to what is the source of such a weak interaction.

In this paper we perform an almost complete characterization on Ca$_3^{2+}$Bi$^{3-}$N$^{3-}$, Sr$_3^{2+}$Bi$^{3-}$N$^{3-}$ and Ba$_3^{2+}$Bi$^{3-}$N$^{3-}$ inverse-perovskite compounds to predict their 'hidden' technological properties. We find very interesting traits in these compounds, some of which have never been reported in literature. The novelty in some cases is attributed to the cation and anion transposition unlike in the parent perovskites and also due to the strong covalency as will be explained herein.

However, it is still unclear whether the obtained characteristic are favorable for novelty or not and this calls for further investigation. 

 The remaining sections of this paper are as systematically arranged in the following fashion:
In Sec.~\ref{sec:technicalities} we give in fine details on how our calculations were done.
 In Sec.~\ref{sec:results} we shade light on structural optimization (Sec.~\ref{structural}), Charge analysis (Sec.~\ref{charge analysis}), the electronic band-structure (Sec.~\ref{Bands}), the active orbital distribution (Sec.~\ref{fat_bands}), the mechanical properties (Sec.~\ref{mechanics})  and the Optical Spectrum (Sec.~\ref{Optical}).

 Future perspectives in the $inverse$ perovskites regarding the hidden technological applications and model making for second-principles simulations are explained in Sec.~\ref{sec:conclusions}.

\section{THE SIMULATION BOX}
 \label{sec:technicalities}
 It is in our expectation that the  semi-local density functionals will always suffer from the self-interaction error~\cite{sic}, which can be avoided by using hybrid density-functional with non-local Hartree-Fock-type exact exchange. Such hybrid density-functional calculations show the great accuracy performance for the geometrical parameters, electronic band structure properties and formation energies of a wide range of materials as reported in Ref.~\onlinecite{Hybrids}. Although such an approach has been highly recommended, it should be taken with a pinch of salt since it forgets some other key interactions like spin-orbit coupling. In such a scenario, it is wise to say that the employment of the HSE06 scheme, is simply an improved LDA+U~\cite{LDA+U}.

 The exchange and correlation functional used in this work was the Generalized Gradient Approximation~\cite{GGA-1996} (GGA-PBE) as embedded in  density functional theory (DFT)~\cite{Hohenberg-64,Kohn-65}. To escape the self interaction error~\cite{DFT_Bandgap_problem} in our materials under study we employed the Heyd-Scuseria-Ernzerhof (HSE06~\cite{HSE06}) screened hybrid functional as embedded in the {\sc Siesta} Method~\cite{SIESTA,New_SIESTA}.

 We replaced the inner electrons with the norm conserving pseudopotentials, made following the Troullier-Martins norm-conserving approach explained in Ref.~\onlinecite{Troullier-91} in the Kleinman-Bylander fully non-local separable representation~\cite{Kleinman-82}.
 During our simulation, we observed that there was an overlap between the semi-core and valence states in Bi, and  thus considering Bi $5d$ electrons into the valence and employing them in the simulation box.
 For reproducibility purposes, finer details regarding the pseudopotentials used in this paper can easily be obtained from Ref.~\onlinecite{Junquera-03.2} for Sr and Ba, Ref.~\onlinecite{Chege} for Bi, Ref.~\onlinecite{ningi2020interplay} for Ca and Ref.~\onlinecite{sifuna2020ab} for N.  Using a recipe found in Ref.~\onlinecite{Sankey-89} and Ref.~\onlinecite{Artacho-99}, we were able to expand the one-electron Kohn-Sham eigenstates in localized numerical atomic orbitals basis set.

 For the special constrain in Bi atom, we employed a single-$\zeta$ basis set for its semi-core states while we used a double-$\zeta$ plus polarization for the valence states for all the atoms herein.

To compute the Hartree, and exchange-potentials, the electronic density and the respective matrix elements between the orbitals, we employed a uniform real-space grid with an equivalent plane-wave cut off of 600 Ry in representation of charge density. 

We employed the Monkhorst-Pack sampling~\cite{Monkhorst-76} for the Brillouin zone integrations equivalent to $11 \times 11 \times 11$ in a five atom $inverse$-perovskite unit cell.
A Fermi-Dirac smearing was chosen for the occupation of the well known one-particle Kohn-Sham electronic eigenstates, with a smearing temperature equalling 0.075 eV.
It is important to note that no constraints were employed in this study apart from the one used in the generation of the Bi pseudopotential. That implies, we allowed both atomic positions and lattice parameters to fully relax.
Starting from such an arrangement, we performed a conjugate gradient minimization until that point when the forces became smaller than $0.01$ eV/\AA ~,and the stress tensor components were below 0.0001 eV/\AA$^3$.

 With the relaxed atomic structure and with a well converged one-particle density matrix, were able to compute the density of states by provoking a non-self consistent calculation now with a denser sampling of $70 \times 70 \times 70$ Monkhorst-Pack mesh.


 Computation of the optical capabilities in the selected $inverse$-perovskites, was done by assuming an external electric field polarized as a mean of three spatial directions, $x,y,z$. The complex dielectric function took the form:
\begin{align}
 \varepsilon(\omega)=\varepsilon_1(\omega)+i\varepsilon_2(\omega),
 \end{align}
 where $\varepsilon_1$ and $\varepsilon_2$ were real and imaginary parts of the dielectric function in that order. Using the recipe described in Ref.~\onlinecite{SIESTA} on the Fermi's golden rule through the inter-bands transitions, we were able to calculate the imaginary part of the complex dielectric function $\varepsilon_2$ as follows:
 \begin{align}
 \varepsilon_2=\frac{4\pi^2}{\Omega \omega^2}\sum_{i\epsilon VB, j\epsilon CB} \sum_{k}W_k\mid \rho _{ij}\mid^2 \delta(\varepsilon_{kj}  -\varepsilon_{ki} -\omega),
 \end{align}
 
 where $W_k$ is the individual $k$-point weight, $\rho_{ij}$ is the dipole transition operator usually projected on the atomic orbitals basis with elements $i$ and $j$. $\Omega$ on the other hand is the unit cell volume, $\omega$, the photon frequency while VB and CB refer to the valence and conduction bands in that order.
 After that is done, we were able to calculate the real part of the complex dielectric function from the Kramers-Kronig notation as explained in Ref.~\onlinecite{TROMER2021138210},

\begin{align}
\varepsilon_1 (\omega)=1+\frac{1}{\pi}P \int_0^\infty \frac{\omega ^\prime \varepsilon_2 (\omega ^\prime)}{\omega ^{\prime 2}-\omega^2}d\omega^\prime,\\
 \end{align}
 where $P$ denotes the principle value. To calculate all the other optical quantities of interest such as the absorption coefficient $\alpha$, the reflectivity $R$, the refractive index $n$ and the extinction coefficient $k$, we employed the following computations: 
 \begin{align}
 &\alpha (\omega)=\sqrt{2}\omega\left[\left(\varepsilon_1^2(\omega)+\varepsilon_2^2(\omega)\right)^{0.5}-\varepsilon_1(\omega)\right]^{0.5},
 \end{align}
 \begin{align}
&R(\omega)=\left[\frac{\left(\varepsilon_1(\omega)+i\varepsilon_2(\omega)\right)^{0.5}-1}{\left(\varepsilon_1(\omega)+i\varepsilon_2(\omega)\right)^{0.5}+1}\right]^{2},
 \label{eqn6}
 \end{align}
 \begin{align}
 &n(\omega)=\frac{1}{\sqrt{2}}\left[\left(\varepsilon_1^2(\omega)+\varepsilon_2^2(\omega)\right)^{0.5}+\varepsilon_1(\omega)\right]^2,
 \end{align}
 \begin{align}
 &k(\omega)=\frac{1}{\sqrt{2}}\left[\left(\varepsilon_1^2(\omega)+\varepsilon_2^2(\omega)\right)^{0.5}-\varepsilon_1(\omega)\right]^2.
 \end{align}
 At this juncture, it is important to note that the accuracy of any optical calculation is highly dependent to the accuracy of the band-gap value. With the well known DFT band gap underestimation, {\sc Siesta} has now implemented the hybrid functionals that heal such and that is the scheme we employed.

 It is common knowledge that the internal energy of an elastic solid is affected by a volumetric strain. This energy is expressed as shown in Eq.~\ref{hooks} If the energy is expanded in a Taylor series up to the second-order about the unstrained situation.
\begin{align}
    E(V,\{\epsilon_k\}) = E_0 + V_0\left(\sum^{6}_{i=1}\sigma_i\epsilon_i + \frac{1}{2}\sum^{6}_{ij=1}C_{ij}\epsilon_{i}\epsilon_{j}\right),
    \label{hooks}
\end{align}
where $\epsilon_k$ denotes $\epsilon_1, \epsilon_2, \cdots, \epsilon_6, V_0(V)$ is the volume, $E_0(E)$ is the energy of an unstrained (strained) system under study while $C_{ij}$ are the elastic constants. 
We calculated the elastic constants by obtaining the second-order partial derivatives of Eq.~\ref{hooks} for both volumetric and distortional deformations
with respect to strains, at zero strains:
\begin{align}
    C_{ij} = \frac{1}{V_0}\left[\frac{\partial^{2}E}{\partial \epsilon_{i}\partial\epsilon_{j}}\right]_{\epsilon_{k=0}},
\label{strains}
\end{align}
In this case, we deformed our crystals by employing the following deformation matrix \textbf{D}:
 \begin{align}
\textbf{D} = 
\begin{pmatrix}
1 + \epsilon & 0 & 0 \\
0 & 1 + \epsilon & 0 \\
0 & 0 & 1 + \epsilon 
\end{pmatrix}.
\end{align}
We then obtained the energy for the above distortions as follows:
\begin{align}
\begin{split}
    E(V, \epsilon) = &E_{0} + V_{0}\epsilon(\sigma_{1} + \sigma_{2} + \sigma_{3}) + \\&V_{0}\left\{\frac{3}{2}(C_{11} + 2C_{12})\epsilon^{2} + O(\epsilon^{3})\right\}.
\end{split}
\end{align}

For us to calculate the born effective charges, we displaced all atoms in the unit cell by a magnitude of 0.01 bohrs along the $x$, $y$ and $z$ directions. The Born effective charge tensor for each atom was then obtained from the change in electric polarization~\cite{king-smith-polarization} triggered by the small displacements on atoms as shown in Eq.~\ref{eq:effective_charge}.
\begin{equation}
    \label{eq:effective_charge}
    Z^*_{i,\alpha,\beta}=\frac{\Omega_0}{e} \left. {\frac{\partial{P_\alpha}}
          {\partial{u_{i,\beta}}}}\right|_{q=0},
  \end{equation}
  where the charge is represented as e and $\Omega_0$ is the volume of the unit cell.

To analyse the fat-bands, we employed a recipe detailed in Ref.~\onlinecite{Chege}, in which the fat-bands ($F_{i,n,\sigma,\vec{k}}$) are said to be periodic equivalent to the Mulliken population.
\begin{align}
    F_{i,n,\sigma,\vec{k}}=\Sigma_j C_{i,n,\sigma,\vec{k}} C_{j,n,\sigma,\vec{k}}  S_{i,j,\vec{k}},
\end{align}
where $C_{i,n,\sigma,\vec{k}}$ are the orbital coefficients while  $S_{i,j,\vec{k}}$ are the overlap matrix elements. $i$ and $j$ denote basis functions while $n$ represents the band index. Conceptualizing the Brillouin zone, $\sigma$ becomes the spin index while $\vec{k}$ is a reciprocal vector.


\section{RESULTS AND DISCUSSIONS}
\label{sec:results}

\subsection{Structural optimization}
\label{structural}
\begin{figure}[H]
	\centering
	\includegraphics [width=\columnwidth]{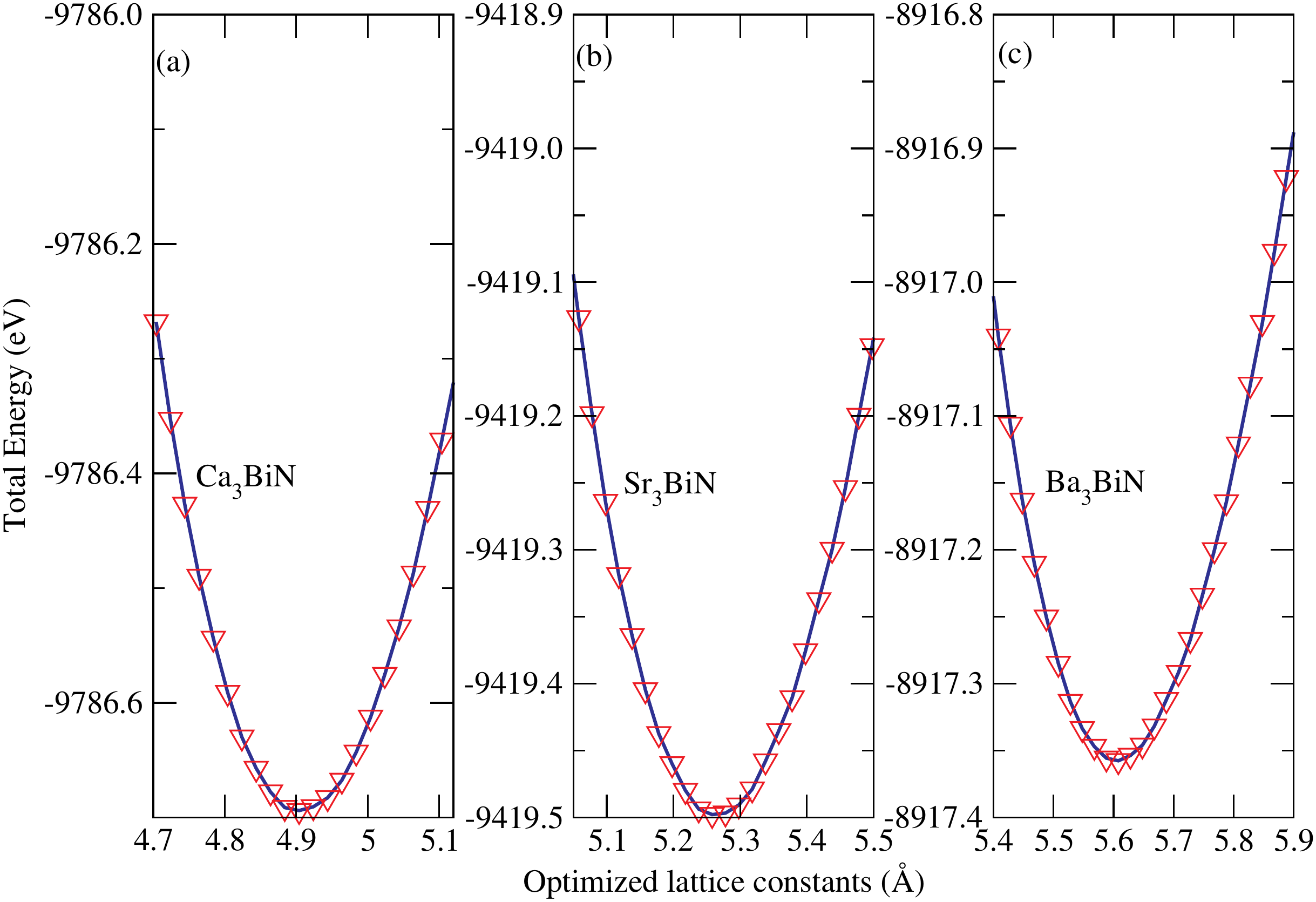}
	\caption {(Color online) A Murnaghan fit~\cite{Murnaghan244} on the selected perovskites $X_3^{2+}$Bi$^{3-}$N$^{3-}$ ($X$=Ca,Sr,Ba). Two things are evident in this figure; (i) the lattice constant of Ba$_3$BiN is larger compared to that of Ca$_3$BiN and Sr$_3$BiN. This is well expected since it is an attribute of the increased atomic radius of barium. (ii) As the lattice constant increases from Ca$_3$BiN $\rightarrow$ Ba$_3$BiN, the total energy in the crystals take a similar trend. This should be taken as such since electrons at higher energy levels depict more energy. Ba$_3$BiN has more high energy electrons unlike Ca$_3$BiN. Some trivial expectations on the three crystals is that Ba$_3$BiN will have the least Bulk modulus due to the long bonds it has. This is indeed confirmed in Table~\ref{table:elastics}.}
	\label{fig:munaghan}
\end{figure}
The calculations on the structural properties of the $inverse$ perovskites were obtained by fitting the unit cell volume versus unit cell energy on the Murnaghan equation as explained in Ref.~\onlinecite{Murnaghan244} of state.
The optimization plots of $X_3^{2+}$Bi$^{3-}$N$^{3-}$ ($X$=Ca,Sr,Ba) is shown in Fig.~\ref{fig:munaghan}.
Based on previous studies~\cite{Low_comp}, it was expected that the generalized gradient approximation employed herein will under-bind the lattice parameters. This was manifested in Table~\ref{table:lattice}. We observed that all the selected $inverse$ perovskites in this study had similar lattice constants just like the traditional perovskites, this is a signature that they will give room to a good match at the interfaces~\cite{interfaces_oxides}.
\begin{table} [H]
	\caption[ ]{ The calculated lattice constants using GGA and HSE06 in \AA.
	}
	\begin{center}
		\begin{tabular}{cccc}
			Crystal	& GGA& HSE06&  EXPT \\
			\hline
			Ca$_3$BiN	&4.90 &4.87  & 4.89~\cite{springer_nitrides} \\
			Sr$_3$BiN	&5.25  &5.20 &  5.21~\cite{Sr3Bi/SbN}\\
			Ba$_3$BiN	&5.59 &5.52 &  -\\
			\hline
		\end{tabular}
	\end{center}
	\label{table:lattice}
\end{table}
Our computed lattice constants are fairly in agreement with available theoretical~\cite{Ullah_2016} and experimental~\cite{springer_nitrides,Sr3Bi/SbN} data. As noted, the value of the lattice constants increase from Ca$_3$BiN$\rightarrow$Sr$_3$BiN$\rightarrow$Ba$_3$BiN because the volume of the unit cell increases as you move down from Ca$\rightarrow$Ba.
\subsection{Charge Analysis}
\label{charge analysis}

\subsubsection{Born effective charge tensors (BEC-T)}
\label{BEC}
We examined the importance of  Born effective charge concept by computing the Born effective charge tensors using a recipe described in Ref.~\onlinecite{Born_effective}. Since our crystals are periodic, we defined the Born effective charge of atom $k$ simply as a coefficient of proportionality at the linear order and under the condition of zero macroscopic electric field, between the macroscopic polarization per unit cell created in the direction  $\beta$ and a cooperative displacement of atoms $k$ in direction $\alpha$~\cite{Ghosez-thesis}:

\begin{align}
Z_{k,\alpha,\beta}^*=\Omega_0\frac{\partial \mathcal{P}_\beta}{\partial \tau_{k,\alpha}}\biggr \rvert_{\varepsilon=0},
\end{align}
where $\Omega_0$ is defined as the unit cell volume.
These charges are important in that they help monitor the long range Coulomb interactions responsible for the splitting between transverse (TO) and longitudinal (LO) optic modes, and governing the optical dielectric constant $\epsilon_\infty$. A more precise observation on Table~\ref{table:BC-values} indicates that the $inverse$-perovskites have '\textit{normal}' $Z^*$ compared to the parent ABO$_3$. The values depicted in Table~\ref{table:BC-values} are slightly equivalent to the nominal charges unlike those of the ABO$_3$ perovskites that at times double in magnitudes.
It is important to note that the acoustic sum rule was preserved by the Born effective charges as shown in Table~\ref{table:BC-values}; that is to say, if we would displace the whole solid rigidly, no polarization will be generated,
\begin{align}
\sum_k Z_{k,\alpha \beta}^*=0.
\end{align} 

\begin{table} [H]
\caption[]{Born effective charges of $X_3^{2+}$Bi$^{3-}$N$^{3-}$ $inverse$-perovskites in their cubic structures. The Born effective charges of the $X$ in comparison to the $nominal$ ionic charges $X$ adapted from Ref.~\onlinecite{Ghosez-thesis}.
}
\begin{center}
\begin{tabular}{ccccccc}
$X_3^{2+}$Bi$^{3-}$N$^{3-}$	& $Z_{Bi}^*$ &$Z_N^*$  & $Z_{X\Vert}^*$ &$Z_{X\perp}^*$  & $Z_{Bi}^*/Z_{Bi}$ &$Z_N^*/Z_N$  \\
	\hline \\
Nominal	& -3.00 & -3.00 & 2.00 &2.00  &  &  \\
Ca$_3$BiN	&-3.23  &-2.92  &0.91  &2.63  & 1.07 &  0.97\\
Sr$_3$BiN	& -3.14 & -2.62 & 0.68 &2.57  & 1.05 & 0.87 \\
Ba$_3$BiN	& -3.28 &-2.34  &0.46  &2.59  & 1.09 &0.78  \\
	\hline
\end{tabular}
	\end{center}
\label{table:BC-values}
\end{table}
In the cubic phase of the $inverse$-perovskites, the Born effective charges are characterized by a set of four independent numbers. The charge on Bi$^{3-}$ and N$^{3-}$ atoms was found to be isotropic due to the spherical symmetry at the atomic sites.  For $X_3^{2+}$ atoms, two independent elements $X\Vert$ and $X\perp$ were considered basing on the atomic displacement parallel and perpendicular to the N$^{3-}$-$X$ bond. We note that the $X$ charge tensor is highly anisotropic. Moreover, the charges on N$^{3-}$ and $X\Vert$ contain additional charges with respect to the nominal ionic value. We equally note that the charge neutrality, reflecting the numerical accuracy of our calculation, is fulfilled to within 0.03. One major striking observation is that the choice of $X_3^{2+}$ has a rather limited influence on $Z_N^*$ and $Z_{X\Vert}^*$ but seems to be dependent on N$^{3-}$ atom.
Another striking observation is that there are additional charges on $Z_N^*$ ) and $Z_{X\perp}^*$. This is due to the more covalent bonding of Bi$^{3-}$ with N$^{3-}$.
\subsubsection{Bader Charges (BC)}
\label{bader}
We define the Bader charges as the charge integrated within a volume. Bader charges are quite different from the Born effective charges. Born effective charges are the electrical polarizations induced by the displacement of the atoms along $x$, $y$ and $z$ directions (that's why it is a tensor and not a scalar parameter). In short, charge analysis involves the net real charge concerning how electrons are transferred or shared between the atoms. In our investigation of the Bader charges in these crystals, we see a strong electron sharing among the atoms. From Table~\ref{table:bader}, we see that the real charges are quite far from the nominal charges and this is a signature of covalency. From Table~\ref{table:bader}, we see that the three compounds have a similar trend. Bi and N have almost the same charge and they portray accumulation of charge (negative charge) while Ca, Sr and Ba show depletion of charge (positive charge). In principle, the real charge in all the atoms increases as we move from Ca$_3$BiN$\rightarrow$Sr$_3$BiN$\rightarrow$Ca$_3$BiN, this is attributed to the decrease in electronegativity down the group. It is wise enough to see charge as a static parameter, and Born effective charges as a response parameter.

\begin{table} [H]
	\caption[ ]{ Calculated Bader charges in reference to the nominal charges for $X_3^{2+}$Bi$^{3-}$N$^{3-}$ ($X$=Ca,Sr,Ba). $Z-N$ is the real charge on a given atom while $Z$ is the number of electrons considered in the valence for the given atom.
	}
	\begin{center}
		\begin{tabular}{cccccccccccccccc}
	
				& & $X$&Bi&N \\
			\hline
			\hline
		Ca$_3$BiN&	Z-N	& 1.29&-1.94&-1.94 \\
		&$nominal$&2.00&-3.00&-3.00\\
		&Z&8.00&15.00&5.00\\
		\hline
		\hline
		Sr$_3$BiN&	Z-N	& 1.27&-1.89&-1.93 \\
		&$nominal$&2.00&-3.00&-3.00\\
		&Z&10.00&15.00&5.00\\
		\hline
			\hline
				Ba$_3$BiN&	Z-N	& 1.16&-1.73&-1.75 \\
		&$nominal$&2.00&-3.00&-3.00\\
		&Z&10.00&15.00&5.00\\
		\hline
			\hline
		\end{tabular}
	\end{center}
	\label{table:bader}
\end{table}

\subsection{Electronic Band-structure}
\label{Bands}
The calculated band-structures  herein imply a semi-conducting trait in the crystals under study with direct gaps at the $\Gamma$ point. The calculations for the band structure reveal a close agreement with other calculated results~\cite{Ullah_2016}.
The calculated results for HSE06 in Table~\ref{table:bandgaps} are better than those of their GGA counter parts although expensive to obtain. One interesting observation from Table~\ref{table:bandgaps} is that the bandgap decreases as the cation changes from Ca$\rightarrow$Sr$\rightarrow$Ba. This decrease of the band gap is due to the decreasing electronegativity of the cations as we go down the group and also due to the increased volume of the unit cell.
\begin{table} [H]
	\caption[ ]{ A comparison of the calculated electronic gaps (eV) using GGA, HSE06 in comparison to the experimental gaps.
	}
	\begin{center}
		\begin{tabular}{cccc}
			Crystal	& GGA& HSE06&  EXPT \\
			\hline
			Ca$_3$BiN	& 0.30&0.99  & 0.95\cite{springer_nitrides} \\
			Sr$_3$BiN	&  0.15&0.94 & 0.89\cite{Rahman_2019} \\
			Ba$_3$BiN	& metalic&0.35 & - \\
			\hline
		\end{tabular}
	\end{center}
	\label{table:bandgaps}
\end{table}

\begin{figure}[H]
\centering
\begin{subfigure}[b]{\columnwidth}
  \includegraphics[width=1\linewidth]{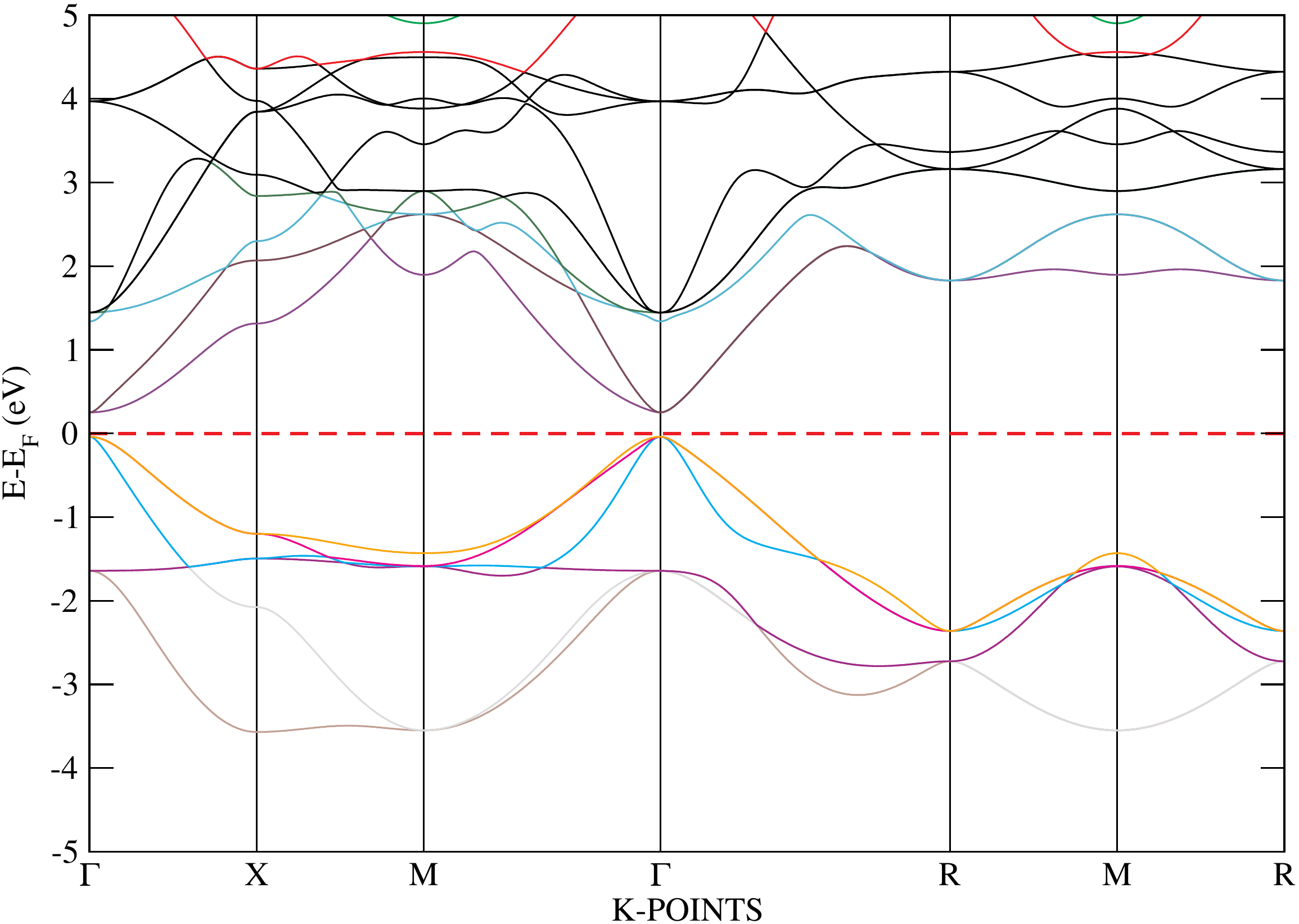}
  \caption{}
  \label{fig:Bands_Ca} 
\end{subfigure}

\begin{subfigure}[b]{\columnwidth}
  \includegraphics[width=1\linewidth]{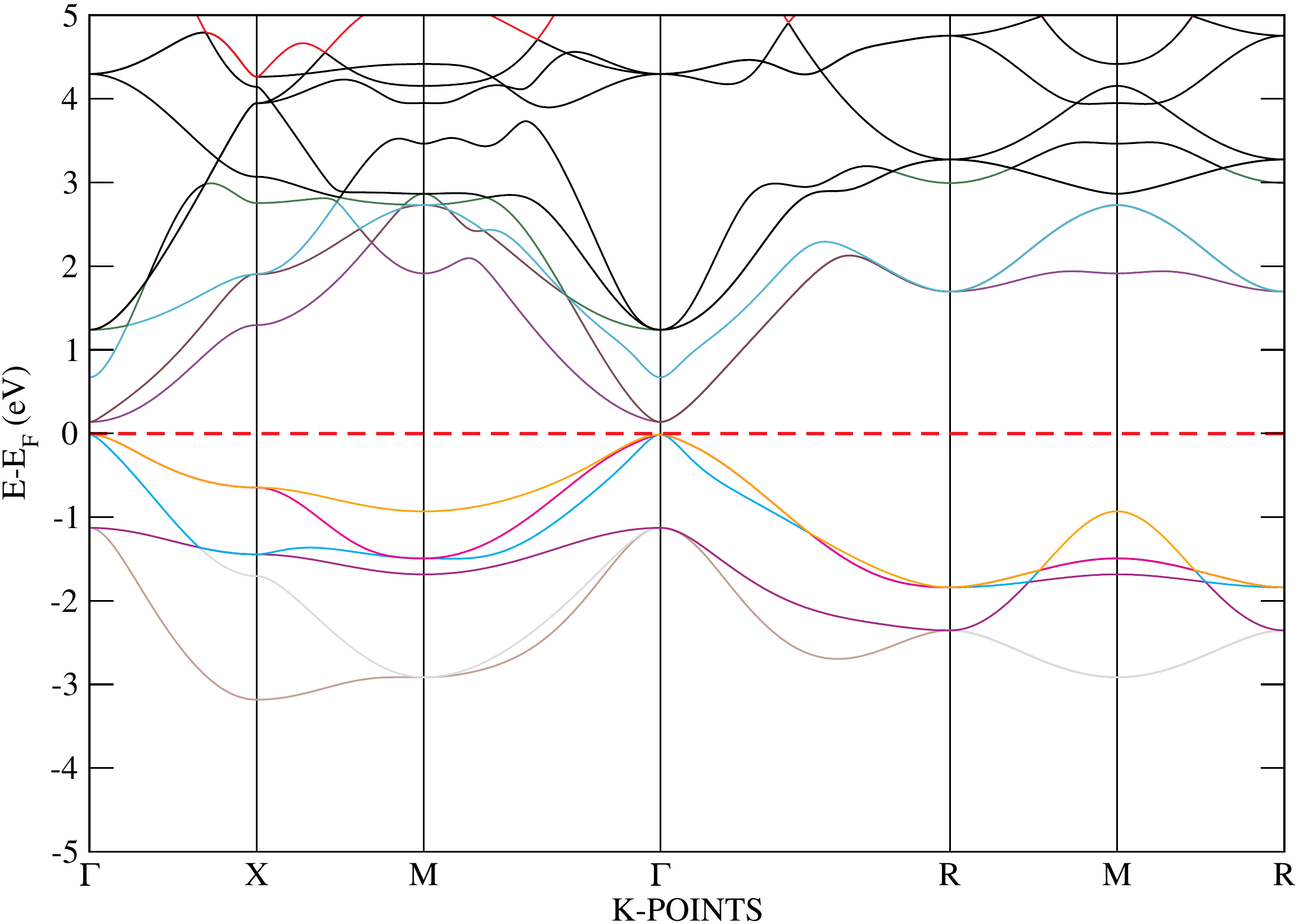}
  \caption{}
  \label{fig:Bands_Sr}
\end{subfigure}

\begin{subfigure}[b]{\columnwidth}
  \includegraphics[width=1\linewidth]{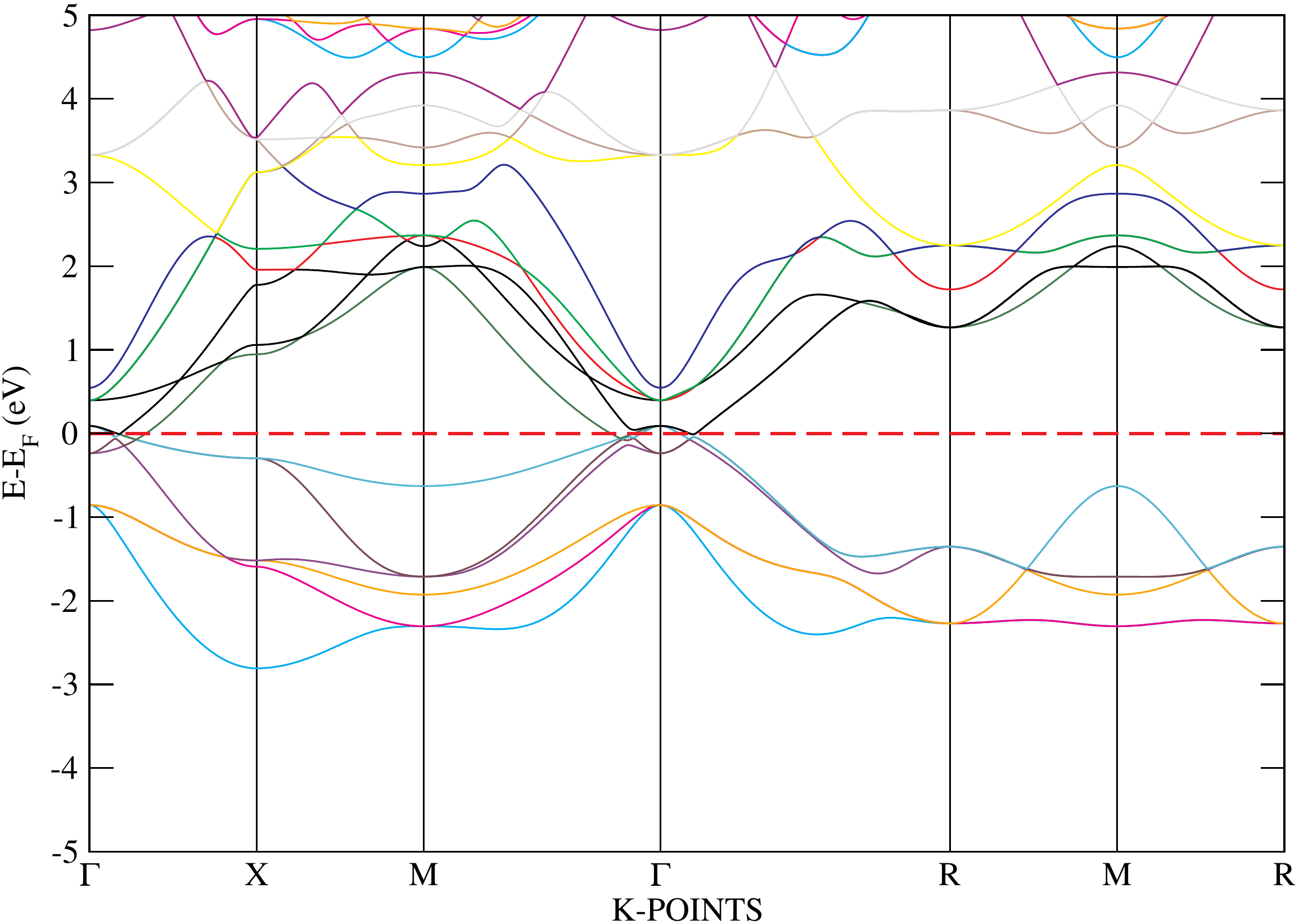}
  \caption{}
  \label{fig:Bands_Ba}
\end{subfigure}
\caption {(Color online) Calculated electronic band structures of $X_3^{2+}$Bi$^{3-}$N$^{3-}$ ($X$=Ca,Sr,Ba) along $\Gamma$-X-M-$\Gamma$-R-M-R high symmetry points. In this case, the Fermi has been set to 0 eV and represented by the red dotted line. (a) Ca$_3$BiN, (b) Sr$_3$BiN and (c) Ba$_3$BiN.}
\label{fig:bandstructure}
\end{figure}

The $inverse$ perovskites under this study have been found to be highly covalent as explained by the charge analysis explained in Sec.~\ref{charge analysis} but also possess a significant ionic
character as shown in Fig.~\ref{fig:bandstructure} and confirmed in Table~\ref{table:bandgaps}. In this case, we have oversimplified the covalency picture in these crystals but it serves well as a starting point for thinking about the electronic properties of this noble materials. In addition to electrostatic interactions, the ions can interact because of the overlap
of the electron wavefunctions. This lead to hybridization between the Bi 6$p$ and N 2$p$
orbitals as depicted in Fig.~\ref{fat_bands} and thus the formation of covalent bonds. Nearly all of the physical and chemical properties in these $inverse$ perovskites are significantly affected by their strong covalency nature as will be explained in the proceeding subsections.
After obtaining the band-structures of the selected $inverse$-perovskites, we went ahead to analyze the character of the different bands to ascertain which atomic orbitals dominated a particular energy window. The science behind the fat band approach is that we are able to plot the well known band eigenvalues and any information regarding the orbital projections on atoms on same footing. In this case, we focused on the top most part (bottom) of the valence band (conduction band) since it is the region where subtle physics occurs.  At the top of the valence band, two mixed

\subsection{Fat-bands description of the active Orbitals}
\label{fat_bands}
\begin{figure}[H]
\centering
\begin{subfigure}[b]{\columnwidth}
  \includegraphics[width=1\linewidth]{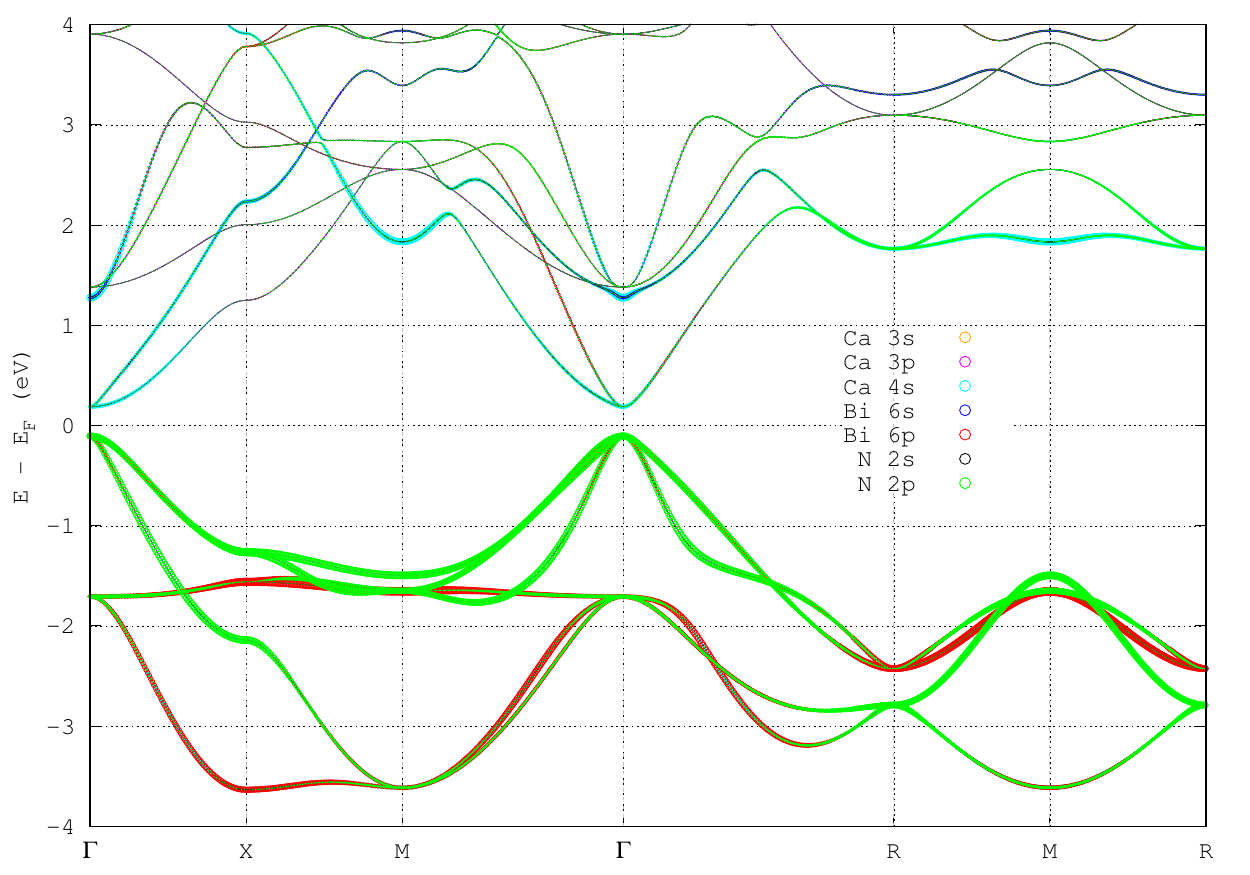}
  \caption{}
  \label{fig:1fats_Ca} 
\end{subfigure}

\begin{subfigure}[b]{\columnwidth}
  \includegraphics[width=1\linewidth]{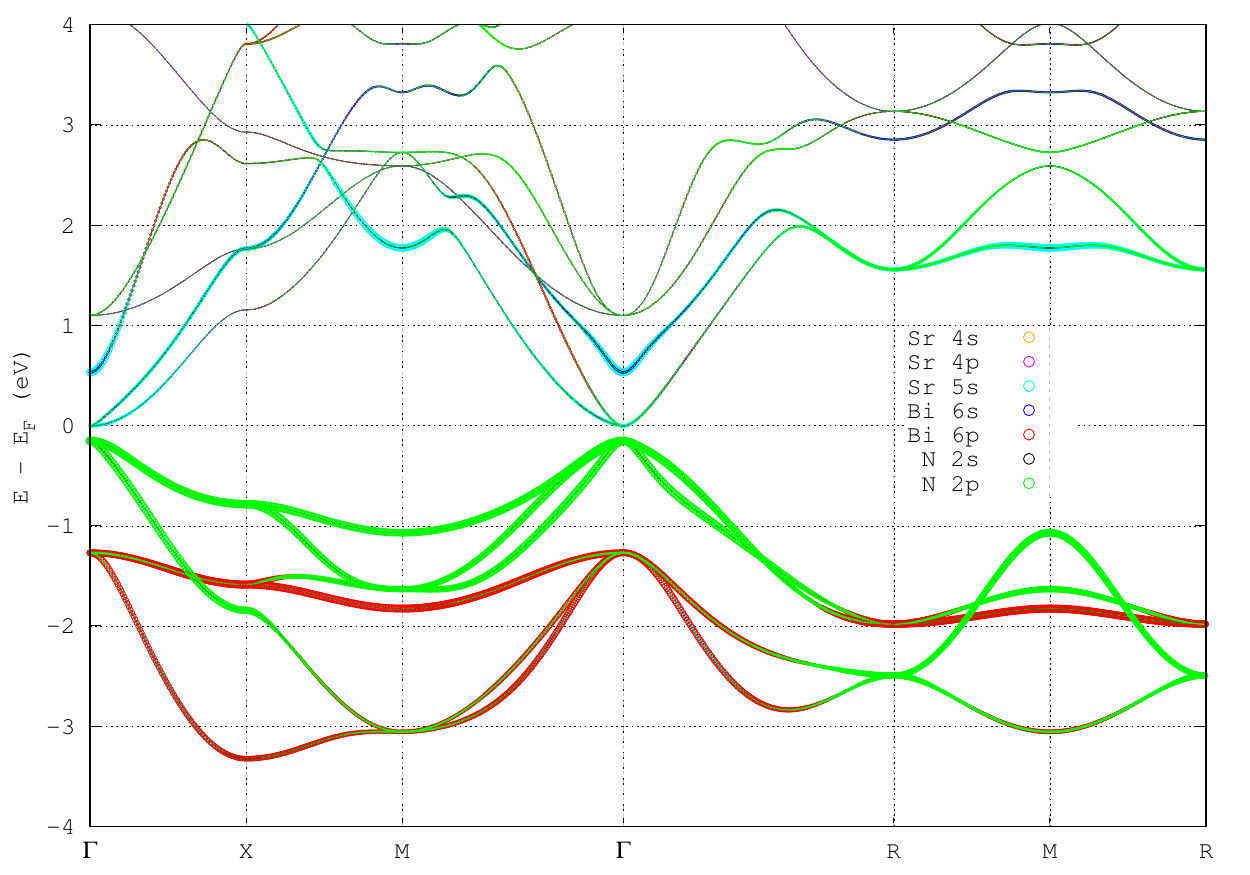}
  \caption{}
  \label{fig:2fats_Sr}
\end{subfigure}

\begin{subfigure}[b]{\columnwidth}
  \includegraphics[width=1\linewidth]{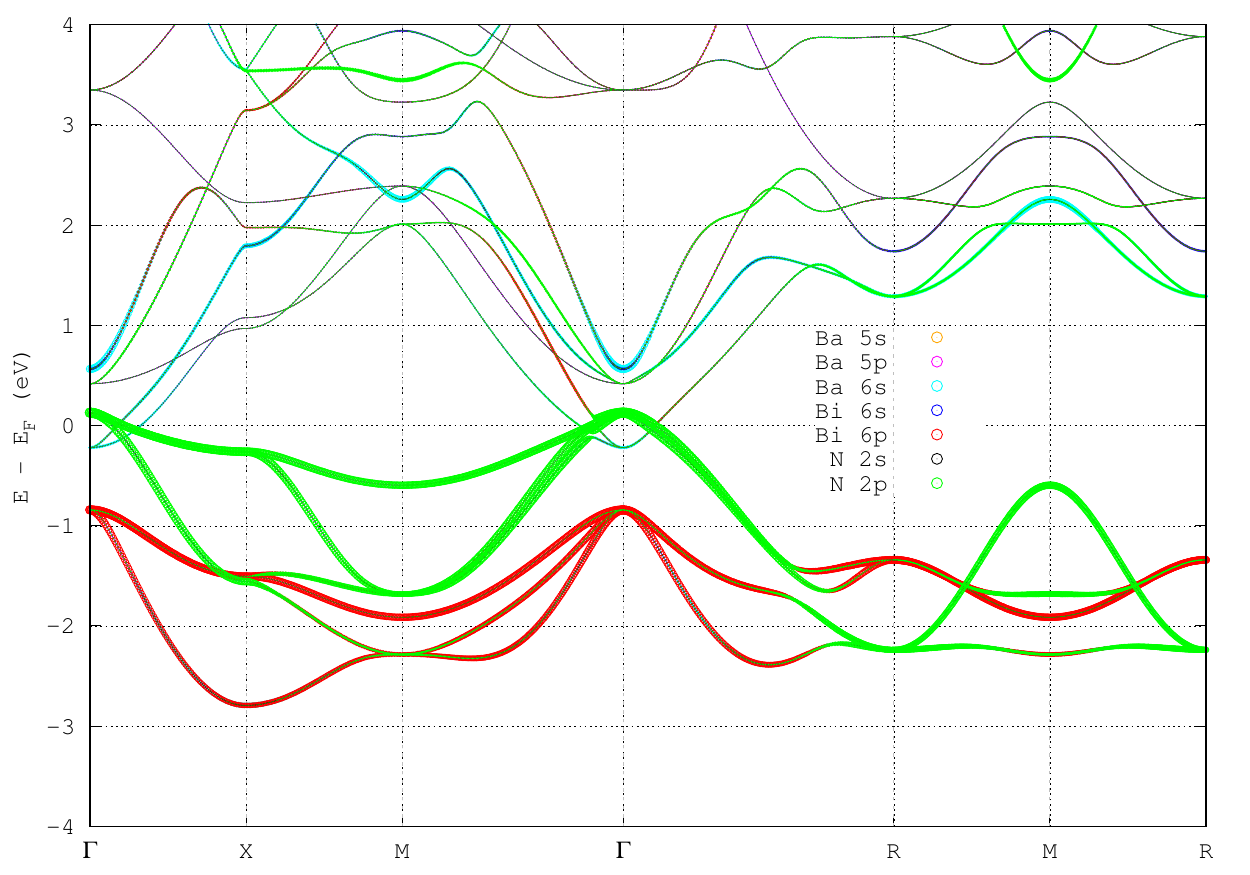}
  \caption{}
  \label{fig:2fats_Ba}
\end{subfigure}
\caption{(Color online)  The band eigenvalues and information about orbital projections on different atoms being treated on the same footing. The bands originating from orbitals deep in energy have been omitted for better clarity of the diagram.}
\end{figure}

structures are observed which are primarily due to the N ($p$) state and Bi ($p$)state. At the bottom of the conduction band, we observed a mixed character. The $s$ orbitals of Ca, Sr and Ba together with the $p$ orbitals of Bi contribute significantly in this window. Also, it is much likely that an admixture is expected between the $p$ state of Bi and N with $s$ state of the Ca, Sr, Ba and thus prompting covalency.
 
\begin{figure*}
	\includegraphics[width=0.8 \columnwidth]{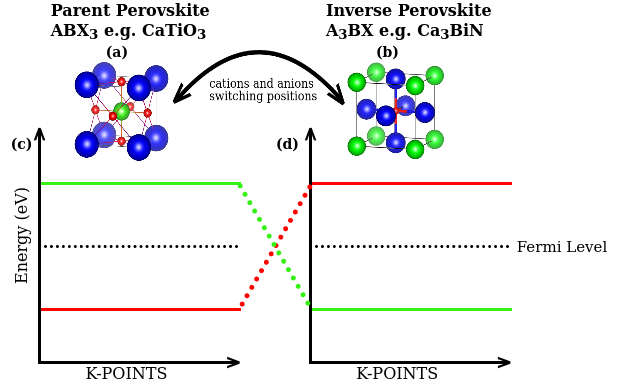}
	\caption {(Color online) (a) and (b), shows the switching of positions between the cations and anions of the parent perovskite and the $inverse$-perovskite. (b) and (c) shows the behaviour of the top and the bottom of the valence band in the parent perovskite and the $inverse$-perovskite respectively.}
	\label{fig:cartoon-bands}
\end{figure*}

We performed a further analysis on the bandstructure in Fig.~\ref{fig:bandstructure} with a comparison with the prototypical ABX$_3$ bandstructure in Fig.~\ref{fig:cartoon-bands}c. From Fig.~\ref{fig:cartoon-bands},  In Fig.~\ref{fig:cartoon-bands}a, we see the atom siting at the center most (0.5,0.5,0.5), has its orbitals projected at the bottom of the conduction band in Fig.~\ref{fig:cartoon-bands}c, the green line. This is contrary to the center most atom in Fig.~\ref{fig:cartoon-bands}b. In these $inverse$-perovskite, the center most atom (0.5,0.5,0.5) has its orbitals projected at the top of the valence band depicted by the green line in Fig.~\ref{fig:cartoon-bands}d. The same analogy is seen on the atoms on the faces of the parent perovskite in Fig.~\ref{fig:cartoon-bands}a, that have their orbitals projected at the top of the valence band as shown by the red line in Fig.~\ref{fig:cartoon-bands}c. Again, this is contrary to the face atoms in Fig.~\ref{fig:cartoon-bands}b, whose projections lie at the bottom of the conduction band as shown by the red line in Fig.~\ref{fig:cartoon-bands}d. Typically, the position switching of the cations and the anions, automatically causes a mirror-like plane on the key projections in the band-structure. This is the reason why we call Fig.~\ref{fig:cartoon-bands}b $inverse$-perovskites. 

\subsection{Mechanical stability and hardness}
\label{mechanics}
Employing the elastic constants, we were able  to predict  Vickers hardness (H$_\nu$) of these noble $inverse$ perovskites using the approach employed in Ref.~\onlinecite{MUCHIRI2019489}. Muchiri and co-workers employed an empirical scheme~\cite{vickers_hardness} that correlates the Vickers hardness to the Pugh's modulus ratio as shown in Eq.~\ref{vickers} below,
\begin{align}
H_\nu=2(k^2G)^{0.585}-3,
\label{vickers}
\end{align}
where $k=\frac{G}{B}$ is the pugh's ratio  while G and B are the shear and bulk modulus respectively.
\begin{table} [H]
	\caption[ ]{ Calculated elastic constants (C$_{11}$, C$_{12}$ and C$_{44}$), B$_0$, $E$ and $G$ in GPa,  the poisson ratio ($\eta$) the density $\rho$ in $g/cm^3$ .
	}
	\begin{center}
		\begin{tabular}{cccccccccccccccc}
			Crystal	& C$_{11}$& C$_{12}$& C$_{44}$&B$_0$&E&G&$\eta$&$\rho$& H$_v$\\
			\hline
			Ca$_3$BiN	& 111.28&23.39  &48.79&52.69&108.31&46.78&0.157&4.8 &13.51 \\
			Sr$_3$BiN	&90.76&17.90&36.57&42.19&85.03&36.52&0.164&5.6 &10.87  \\
			Ba$_3$BiN	&70.07&16.98&25.61&34.67&62.36&25.98&0.200&6.1 &8.36  \\
			\hline
		\end{tabular}
	\end{center}
	\label{table:elastics}
\end{table}
From the computations, we can see that $C_{11}$ for Ca$_3$BiN is higher than that of the other two which signifies a larger stiffness against strain. Checking at the $C_{44}$ value that is also larger in Ca$_3$BiN, we conclusively state that that of the three, Ba$_3$BiN has the least resistance to shear strain. Such kind of comparison should be taken as such since it is well expected. Ba$_3$BiN has longer bonds than those found in Ca$_3$BiN. To check at the mechanical stability of these crystals, we employed the Born criteria~\cite{born_1940} which  consists of the four listed equations equations below.  
\begin {align}
C_{11}-C_{12}>0\\
C_{11}>0\\
C_{44}>0\\
(C_{11}+2C_{12})>0
\end{align}
From our calculations, it is evident that the Born's stability criteria is satisfied and it verifies the mechanical stability of cubic Ca$_3$BiN, Sr$_3$BiN and Ba$_3$BiN crystal structures. To further analyse our mechanical calculations, we employed the Voigt-Ress-Hill~\cite{Hill} mechanical constants to calculate the bulk modulus ($B$), Young’s modulus ($E$), shear modulus ($G$) and Poisson’s ratio $\eta$.
\begin{align}
&B=\frac{1}{3}(C_{11}+2C_{12})\\
&G_V=\frac{1}{5}(C_{11}-C_{12}+3C_{44})\\
&G_R=\frac{5(C_{11}-C_{12})C_{44}}{4C_{44}+3(C_{11}-C_{12})}\\
&G_H=\frac{1}{2}(G_V+G_R)\\
&E=\frac{9BG_H}{3B+G_H}\\
&\eta=\frac{3B-2G_H}{2(3B+G_H)}
\end{align}

Bulk modulus ($B$) reflects the strength of a material. In $X_3^{2+}$Bi$^{3-}$N$^{3-}$ ($X$=Ca,Sr,Ba) compounds, we can arrange the bulk modulus as follows; Ca$_3$BiN $>$ Sr$_3$BiN $>$ Ba$_3$BiN.  The bulk modulus trend is typically of an inverse sequence to the trend seen in the lattice constants in Table~\ref{table:lattice}, so it is decreasing down the group. The Shear modulus ($G$) characterizes a materials resistance to shear strain. Again, the order is similar to that of the Bulk modulus ($B$) and the attribution maintained. The Young's modulus ($E$) measures a materials stiffness. An observation on Table~\ref{table:elastics}, we see clearly that Ca$_3$BiN is the stiffest among the three crystals under study. 
According to Pugh's theory, the ration $\frac{B}{G}$ is strongly related to a materials ductility. If $\frac{B}{G}$ is greater than 1.75, then the material exhibits ductility, otherwise, the material exhibits brittleness. A close look on Table~\ref{table:elastics} shows a plastic nature of the three crystals under study. They will easily fracture with little elastic deformation. This observation is confirmed by the Poisson's ratio ($\eta$) in the sense that all our crystals herein have $\eta<0.33$ implying a strong covalent nature. 
\subsection{Response to light}

\label{Optical}
Before embarking on this section, it is wise to say something on our approach. When using the {\sc Siesta} method to calculate the optical response in materials, we expect a deviation at higher energies since the quality of the conduction-band states is worse than that of the valence-band ones. It is also significant to note that in the {\sc Siesta} method, the basis sets we employed were relatively smaller than in other codes (typically plane-wave), and that left us with a limited number of bands to employ. Either way, the output from our work is comparable to the output from other methods and the differences are not significative at all. 

In this section we discuss the optical parameters of the $X_3^{2+}$Bi$^{3-}$N$^{3-}$ ($X$=Ca,Sr,Ba) compounds. Specifically, we explore the frequency dependent real $\varepsilon_1$ ($\omega$) , and imaginary $\varepsilon_2$ ($\omega$) parts of the complex dielectric function. We also investigate the refractive index n($\omega$), extinction coefficient k($\omega$), absorption coefficient $\alpha$($\omega$) and reflectivity R($\omega$)  of the compounds. 
The dispersion of linear optical response of the title materials are presented in the optical window spanning from 0 to 15 eV. This highlights the response from the static limit to ultraviolet regime of the electromagnetic spectrum. 
\subsubsection{Real and Imaginary parts of the dielectric constant in  \texorpdfstring{$X_3^{2+}Bi^{3-}N^{3-}$} (X=Ca,Sr,Ba) inverse-perovskites}
\begin{figure}[H]
	\centering
	\includegraphics [width=\columnwidth]{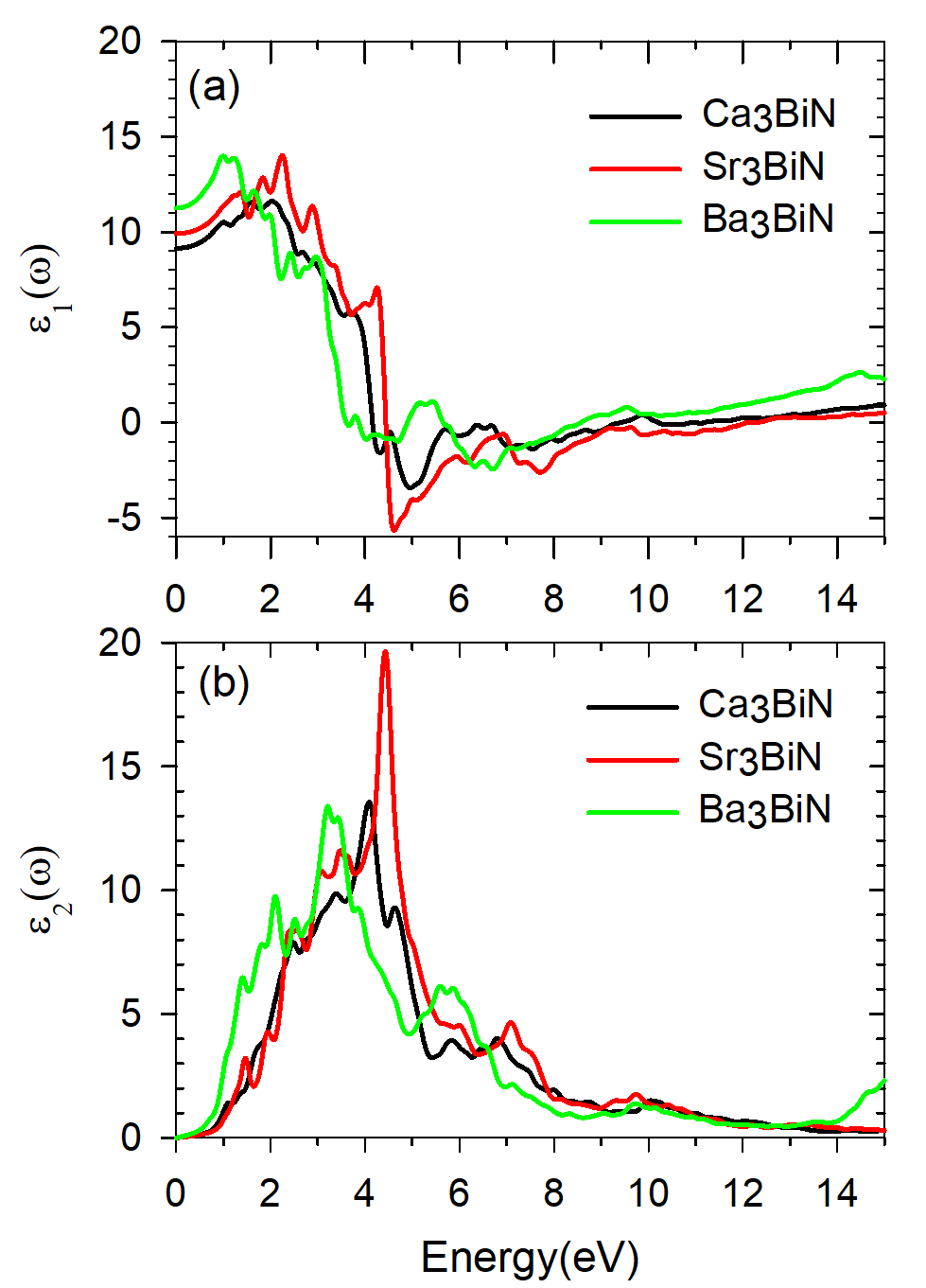}
	\caption {(Color online)The real and imaginary parts of the dielectric constant as a function of energy for $X_3^{2+}$Bi$^{3-}$N$^{3-}$ ($X$=Ca,Sr,Ba) $inverse$-perovskites.}
	\label{fig:real_imag}
\end{figure}

Fig.~\ref{fig:real_imag}a and Fig.~\ref{fig:real_imag}b shows the real and imaginary parts of the dielectric function, respectively. The static limit real part of the dielectric function for Ca$_3$BiN, Sr$_3$BiN, and Ba$_3$BiN are 9.1434, 9.9135 and 11.2508, respectively. This corresponds to the zero frequency, $\varepsilon_1$ (0), response of the materials. The static limit real part of the dielectric function $\varepsilon_1$ (0)(with the imaginary part $\varepsilon_2$ (0) being zero) has an empirical inverse dependence with bandgap of the material under study, this is commonly known as Moss relation~\cite{TMOSS}. As the bandgap of the material decreases the corresponding $\varepsilon_1$ (0) is expected to increase which is in agreement with the reported values presented here. The maximum values of the real part of the dielectric function for Ca$_3$BiN, Sr$_3$BiN, and Ba$_3$BiN are 11.6204, 14.0234 and 13.9910 occur at the energies of 2.0420 eV, 2.2422 eV and 1.0010 eV respectively. The dielectric function crosses over to the negative values and further shifts to positive at higher energies. For the energy ranges when $\varepsilon_1$ ($\omega$) is negative the materials may be viewed as behaving as a metal. This metallic behavior is attributed to excitation of electrons from the valence band to conduction band.
Fig.~\ref{fig:real_imag}b shows the dispersion of the imaginary part of the dielectric function. Notable peaks can be observed in the optical spectra of the compounds in line with inter-band transitions of electrons from the valence band to conduction band. The maximum values of the imaginary part of the dielectric function for Ca$_3$BiN, Sr$_3$BiN, and Ba$_3$BiN are 13.5679, 19.5990 and 13.3950 at the energies of 4.0841 eV, 4.4444 eV and 3.2032 eV respectively. These peaks are within the ultraviolet regime and hints to strong optical response within the range.

\subsubsection{Refractive index in  \texorpdfstring{$X_3^{2+}$Bi$^{3-}$N$^{3-}$} (X=Ca,Sr,Ba) inverse-perovskites}
\begin{figure}[H]
	\centering
	\includegraphics [width=\columnwidth]{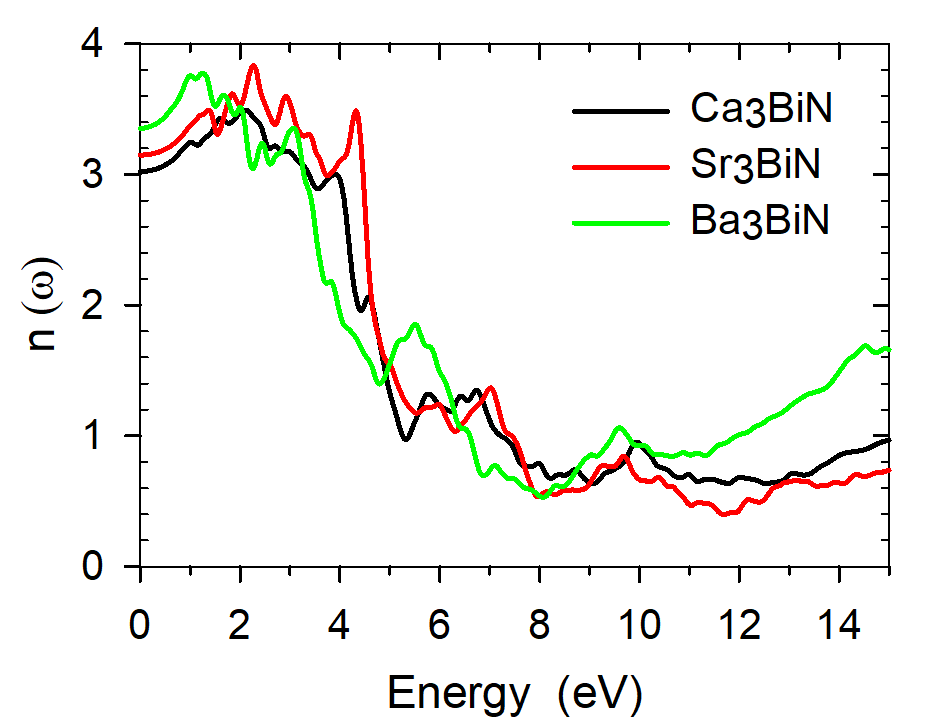}
	\caption {(Color online) The refractive index as a function of energy for $X_3^{2+}$Bi$^{3-}$N$^{3-}$ ($X$=Ca,Sr,Ba) $inverse$-perovskites.}
	\label{fig:refractive_indx}
\end{figure}

The dispersion of the refractive index n($\omega$) is represented in Fig.~\ref{fig:refractive_indx}. For lower energies the refractive index monotonically increases and attains the maximum value as the energy increases. The static limit refractive index n(0) for the Ca$_3$BiN, Sr$_3$BiN, and Ba$_3$BiN are 3.024, 3.149 and 3.354, respectively. As mentioned earlier, the Moss relation~\cite{TMOSS} can be verified since static limit refractive index n(0) has inverse power law dependence with the bandgap. An increase in n(0) corresponds to decrease in bandgap energy.  Meanwhile, the peak values of the refractive indices lies in the energy range of 1.2 eV to 2.3 eV. The maximum values of the refractive indices was found to be 3.488, 3.827 and 3.772 at 2.082 eV, 2.242 eV, and 1.281 eV for Ca$_3$BiN, Sr$_3$BiN, and Ba$_3$BiN respectively as shown Table~\ref{table:refractive_ext}.

\subsubsection{Extinction coefficient in \texorpdfstring{$X_3^{2+}$Bi$^{3-}$N$^{3-}$} (X=Ca,Sr,Ba) inverse-perovskites}
\begin{figure}[H]
	\centering
	\includegraphics [width=\columnwidth]{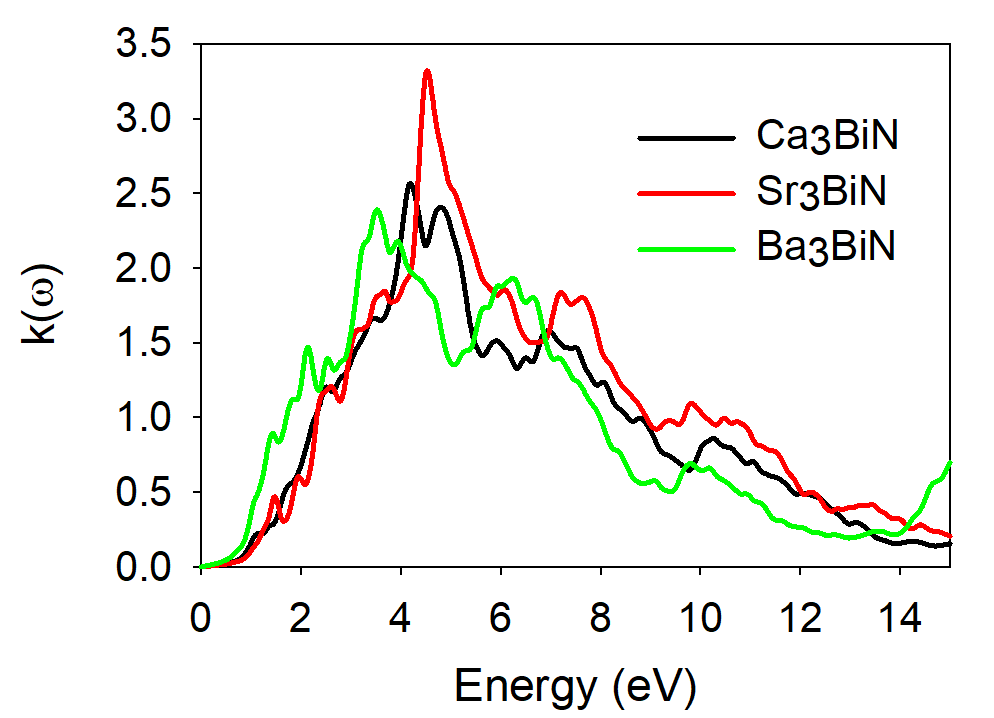}
	\caption {(Color online) The extinction coefficient as a function of energy for $X_3^{2+}$Bi$^{3-}$N$^{3-}$ ($X$=Ca,Sr,Ba) $inverse$-perovskites.}
	\label{fig:extinction_coeff}
\end{figure}
Fig.~\ref{fig:extinction_coeff} shows the dispersion relation for the extinction coefficient k ($\omega$).  The extinction coefficient modulates in the same fashion as the imaginary part of the dielectric function due to their direct relationship,
\begin{align}
\alpha=\frac{2k\omega}{c}=\frac{4(\pi)k}{\lambda},
\end{align}
where $c$ is the speed of light and $\lambda$ is the the wavelength.
The peak values of the extinction coefficient are given in Table~\ref{table:refractive_ext}.

\begin{table} [H]
	\caption[ ]{Refractive index and extinction coefficient}
	\begin{center}
\begin{tabular}{ccccccc}
	& n (0) &n ($\omega$)$_{max}$ &k ($\omega$)$_{max}$\\
	&&(at energy (eV))&(at energy (eV))\\
	\hline \\
Ca$_3$BiN	&3.024  &3.488 (2.082)  &2.5600 (4.164)\\
Sr$_3$BiN& 3.149 & 3.827( 2.242)&3.2920 (4.565) \\
Ba$_3$BiN& 3.354&3.772 (1.281) &2.3900 (3.523) \\
	\hline
\end{tabular}
	\end{center}
\label{table:refractive_ext}
\end{table}

\subsubsection{Reflectivity in \texorpdfstring{$X_3^{2+}$Bi$^{3-}$N$^{3-}$} (X=Ca,Sr,Ba) inverse-perovskites }
\begin{figure}[H]
	\centering
	\includegraphics [width=\columnwidth]{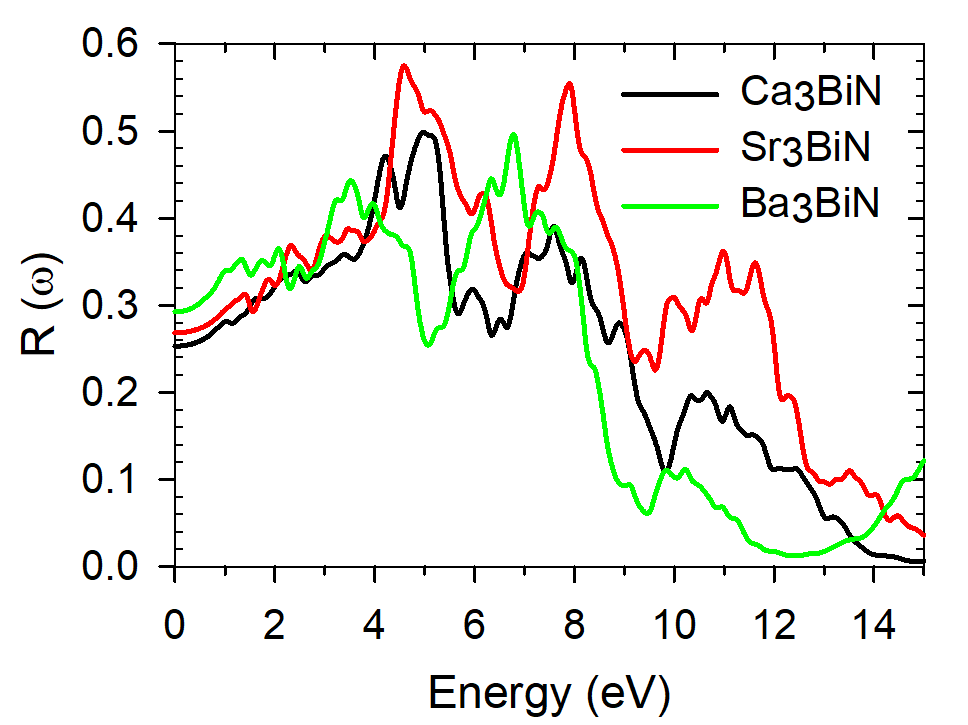}
	\caption {(Color online) Reflectivity as a function of energy for $X_3^{2+}$Bi$^{3-}$N$^{3-}$ ($X$=Ca,Sr,Ba) $inverse$-perovskites.}
	\label{fig:reflectivity}
\end{figure}

The reflection of materials on surfaces is determined by the coefficient of reflection or reflectivity. Reflectivity is defined as the quotient of the reflected power to the incident power on the surface. The frequency dependent reflectivity of our materials was computed based on Eq.~(\ref{eqn6}). Reflectivity of the compounds are plotted in Fig.~\ref{fig:reflectivity}. The zero frequency reflectivity is between 25\% to 29\% for all the materials, and monotonically increases in the energy range from 0 to 1.5 eV. The peak reflectivity values for the Ca$_3$BiN, Sr$_3$BiN, and Ba$_3$BiN are 50\%, 57\%, and 49\% at the energies of 4.965 eV, 4.565 eV and 6.807 eV respectively.

\subsubsection{Absorption coefficient in \texorpdfstring{$X_3^{2+}$Bi$^{3-}$N$^{3-}$} (X=Ca,Sr,Ba) inverse-perovskites }
\begin{figure}[H]
	\centering
	\includegraphics [width=\columnwidth]{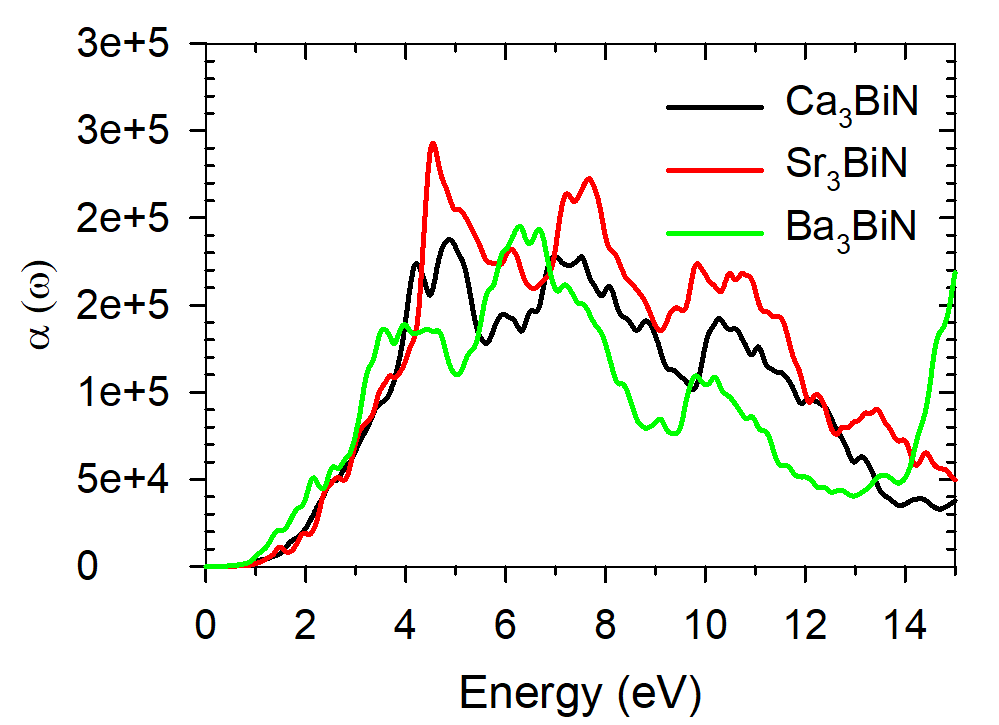}
	\caption {(Color online) Absorption coefficient as a function of energy for $X_3^{2+}$Bi$^{3-}$N$^{3-}$ ($X$=Ca,Sr,Ba) $inverse$-perovskites.}
	\label{fig:absorption}
\end{figure}

The absorption of the incident photon by an optical material is described by the frequency dependent absorption coefficient $\alpha$. The absorption coefficient for $X_3^{2+}$Bi$^{3-}$N$^{3-}$ ($X$=Ca,Sr,Ba) are presented in Fig.~\ref{fig:absorption}. The figure depicts that the coefficient of absorption for the Ca$_3$BiN, Sr$_3$BiN, and Ba$_3$BiN strongly modulates indicating optical transitions within our observation range with maximum peak values of  $1.876 \times 10^5$  cm$^{-1}$, $2.424\times 10^5$ cm$^{-1}$ and $1.946 \times 10^5$ cm$^{-1}$ at energies of 4.885 eV, 4.565 eV and 6.246 eV respectively. The high absorption coefficients indicates that our title materials are good optical absorbers, which enhances promotion of excited electrons from the maximum of the valence band to the conduction band. These materials are promising in applications for optoelectronic devices in the ultraviolet regime. 

\section{CONCLUSION AND FUTURE WORKS}
\label{sec:conclusions}
In this work, we have studied the energetics, band structure, charge analysis, mechanical stability and the optical spectrum of $X_3^{2+}$Bi$^{3-}$N$^{3-}$ ($X$=Ca,Sr,Ba). The calculated lattice match is similar to those of the $ABX_3$ parent perovskites indicating a good lattice match for future interfacing capabilities.
We observe also a very strong covalency emanating from the charge sharing. This is confirmed by from the hybridization in the band-structure and the Poisson's ratio. Mechanically, all the crystals in this study are stable and have a plastic character attached to them. 

The direct band-gaps obtained are in much agreement with the theoretical existing gaps. All the crystals are semiconducting in nature.
The optical spectrum on the other hand is interesting in the sense that all the crystals under the study have a strong optical activity in the ultraviolet regime. 

With good models on these systems, we wonder if second principles calculations~\cite{second_principles} could be employed so as to have both the lattice and electronic parts on the same footing as explained in Ref.~\onlinecite{mrs-advances-james}. The beauty of this will be to see the behaviour of these materials at elevated temperatures. \\

\section*{ACKNOWLEDGMENT} 
\label{sec:Acknowledgement} 
 We thank Yann Pouillon, of Simune Atomistics for providing us with the first ever {\sc Siesta} tarball implemented with the HSE06 formalism. We thank Javier Junquera for his technical assistance in improving the quality of the basis sets used in this study. Mirriam Chepkoech is thanked for the fruitful discussions we had on these novel Crystals. Support from the Centre for High Performance Computing (CHPC-MATS1424), Cape Town, South Africa is appreciated. We acknowledge financial support from Masinde Muliro University of Science and Technology Grant No. MMU/URF/2022/1-026.
 \section*{DATA AVAILABILITY STATEMENT}
\label{sec:data_availability}
 Can be requested from the corresponding author.
\bibliography{james}

\begin{thebibliography}{60}%
\makeatletter
\providecommand \@ifxundefined [1]{%
 \@ifx{#1\undefined}
}%
\providecommand \@ifnum [1]{%
 \ifnum #1\expandafter \@firstoftwo
 \else \expandafter \@secondoftwo
 \fi
}%
\providecommand \@ifx [1]{%
 \ifx #1\expandafter \@firstoftwo
 \else \expandafter \@secondoftwo
 \fi
}%
\providecommand \natexlab [1]{#1}%
\providecommand \enquote  [1]{``#1''}%
\providecommand \bibnamefont  [1]{#1}%
\providecommand \bibfnamefont [1]{#1}%
\providecommand \citenamefont [1]{#1}%
\providecommand \href@noop [0]{\@secondoftwo}%
\providecommand \href [0]{\begingroup \@sanitize@url \@href}%
\providecommand \@href[1]{\@@startlink{#1}\@@href}%
\providecommand \@@href[1]{\endgroup#1\@@endlink}%
\providecommand \@sanitize@url [0]{\catcode `\\12\catcode `\$12\catcode
  `\&12\catcode `\#12\catcode `\^12\catcode `\_12\catcode `\%12\relax}%
\providecommand \@@startlink[1]{}%
\providecommand \@@endlink[0]{}%
\providecommand \url  [0]{\begingroup\@sanitize@url \@url }%
\providecommand \@url [1]{\endgroup\@href {#1}{\urlprefix }}%
\providecommand \urlprefix  [0]{URL }%
\providecommand \Eprint [0]{\href }%
\providecommand \doibase [0]{http://dx.doi.org/}%
\providecommand \selectlanguage [0]{\@gobble}%
\providecommand \bibinfo  [0]{\@secondoftwo}%
\providecommand \bibfield  [0]{\@secondoftwo}%
\providecommand \translation [1]{[#1]}%
\providecommand \BibitemOpen [0]{}%
\providecommand \bibitemStop [0]{}%
\providecommand \bibitemNoStop [0]{.\EOS\space}%
\providecommand \EOS [0]{\spacefactor3000\relax}%
\providecommand \BibitemShut  [1]{\csname bibitem#1\endcsname}%
\let\auto@bib@innerbib\@empty
\bibitem [{\citenamefont {Wang}\ \emph {et~al.}(2020)\citenamefont {Wang},
  \citenamefont {Zhang}, \citenamefont {Zhu}, \citenamefont {Lü},
  \citenamefont {Li}, \citenamefont {Zou},\ and\ \citenamefont
  {Zhao}}]{Anti-perovskites1}%
  \BibitemOpen
  \bibfield  {author} {\bibinfo {author} {\bibfnamefont {Y.}~\bibnamefont
  {Wang}}, \bibinfo {author} {\bibfnamefont {H.}~\bibnamefont {Zhang}},
  \bibinfo {author} {\bibfnamefont {J.}~\bibnamefont {Zhu}}, \bibinfo {author}
  {\bibfnamefont {X.}~\bibnamefont {Lü}}, \bibinfo {author} {\bibfnamefont
  {S.}~\bibnamefont {Li}}, \bibinfo {author} {\bibfnamefont {R.}~\bibnamefont
  {Zou}}, \ and\ \bibinfo {author} {\bibfnamefont {Y.}~\bibnamefont {Zhao}},\
  }\href {\doibase https://doi.org/10.1002/adma.201905007} {\bibfield
  {journal} {\bibinfo  {journal} {Advanced Materials}\ }\textbf {\bibinfo
  {volume} {32}},\ \bibinfo {pages} {1905007} (\bibinfo {year}
  {2020})}\BibitemShut {NoStop}%
\bibitem [{\citenamefont {Ochi}\ and\ \citenamefont
  {Kuroki}(2019{\natexlab{a}})}]{PhysRev}%
  \BibitemOpen
  \bibfield  {author} {\bibinfo {author} {\bibfnamefont {M.}~\bibnamefont
  {Ochi}}\ and\ \bibinfo {author} {\bibfnamefont {K.}~\bibnamefont {Kuroki}},\
  }\href {\doibase 10.1103/PhysRevApplied.12.034009} {\bibfield  {journal}
  {\bibinfo  {journal} {Phys. Rev. Applied}\ }\textbf {\bibinfo {volume}
  {12}},\ \bibinfo {pages} {034009} (\bibinfo {year}
  {2019}{\natexlab{a}})}\BibitemShut {NoStop}%
\bibitem [{\citenamefont {Pertsova}\ \emph {et~al.}(2019)\citenamefont
  {Pertsova}, \citenamefont {Geilhufe}, \citenamefont {Bremholm},\ and\
  \citenamefont {Balatsky}}]{PhysRevB.99.205126}%
  \BibitemOpen
  \bibfield  {author} {\bibinfo {author} {\bibfnamefont {A.}~\bibnamefont
  {Pertsova}}, \bibinfo {author} {\bibfnamefont {R.~M.}\ \bibnamefont
  {Geilhufe}}, \bibinfo {author} {\bibfnamefont {M.}~\bibnamefont {Bremholm}},
  \ and\ \bibinfo {author} {\bibfnamefont {A.~V.}\ \bibnamefont {Balatsky}},\
  }\href {\doibase 10.1103/PhysRevB.99.205126} {\bibfield  {journal} {\bibinfo
  {journal} {Phys. Rev. B}\ }\textbf {\bibinfo {volume} {99}},\ \bibinfo
  {pages} {205126} (\bibinfo {year} {2019})}\BibitemShut {NoStop}%
\bibitem [{\citenamefont {Chepkoech}\ \emph {et~al.}(2020)\citenamefont
  {Chepkoech}, \citenamefont {Joubert},\ and\ \citenamefont
  {Amolo}}]{CHEPKOECH2020e00484}%
  \BibitemOpen
  \bibfield  {author} {\bibinfo {author} {\bibfnamefont {M.}~\bibnamefont
  {Chepkoech}}, \bibinfo {author} {\bibfnamefont {D.~P.}\ \bibnamefont
  {Joubert}}, \ and\ \bibinfo {author} {\bibfnamefont {G.~O.}\ \bibnamefont
  {Amolo}},\ }\href {\doibase https://doi.org/10.1016/j.cocom.2020.e00484}
  {\bibfield  {journal} {\bibinfo  {journal} {Computational Condensed Matter}\
  }\textbf {\bibinfo {volume} {24}},\ \bibinfo {pages} {e00484} (\bibinfo
  {year} {2020})}\BibitemShut {NoStop}%
\bibitem [{\citenamefont {Kariyado}\ and\ \citenamefont {Ogata}(2011)}]{Japan}%
  \BibitemOpen
  \bibfield  {author} {\bibinfo {author} {\bibfnamefont {T.}~\bibnamefont
  {Kariyado}}\ and\ \bibinfo {author} {\bibfnamefont {M.}~\bibnamefont
  {Ogata}},\ }\href {\doibase 10.1143/JPSJ.80.083704} {\bibfield  {journal}
  {\bibinfo  {journal} {Journal of the Physical Society of Japan}\ }\textbf
  {\bibinfo {volume} {80}},\ \bibinfo {pages} {083704} (\bibinfo {year}
  {2011})}\BibitemShut {NoStop}%
\bibitem [{\citenamefont {Kariyado}\ and\ \citenamefont
  {Ogata}(2017)}]{Kariyado_PRB}%
  \BibitemOpen
  \bibfield  {author} {\bibinfo {author} {\bibfnamefont {T.}~\bibnamefont
  {Kariyado}}\ and\ \bibinfo {author} {\bibfnamefont {M.}~\bibnamefont
  {Ogata}},\ }\href {\doibase 10.1103/PhysRevMaterials.1.061201} {\bibfield
  {journal} {\bibinfo  {journal} {Phys. Rev. Materials}\ }\textbf {\bibinfo
  {volume} {1}},\ \bibinfo {pages} {061201} (\bibinfo {year}
  {2017})}\BibitemShut {NoStop}%
\bibitem [{\citenamefont {Kaur}\ and\ \citenamefont
  {Sinha}(2020)}]{KAUR2020101741}%
  \BibitemOpen
  \bibfield  {author} {\bibinfo {author} {\bibfnamefont {T.}~\bibnamefont
  {Kaur}}\ and\ \bibinfo {author} {\bibfnamefont {M.}~\bibnamefont {Sinha}},\
  }\href {\doibase https://doi.org/10.1016/j.mtcomm.2020.101741} {\bibfield
  {journal} {\bibinfo  {journal} {Materials Today Communications}\ ,\ \bibinfo
  {pages} {101741}} (\bibinfo {year} {2020})}\BibitemShut {NoStop}%
\bibitem [{\citenamefont {Nuss}\ \emph {et~al.}(2015)\citenamefont {Nuss},
  \citenamefont {M{\"{u}}hle}, \citenamefont {Hayama}, \citenamefont
  {Abdolazimi},\ and\ \citenamefont {Takagi}}]{Nuss:dk5032}%
  \BibitemOpen
  \bibfield  {author} {\bibinfo {author} {\bibfnamefont {J.}~\bibnamefont
  {Nuss}}, \bibinfo {author} {\bibfnamefont {C.}~\bibnamefont {M{\"{u}}hle}},
  \bibinfo {author} {\bibfnamefont {K.}~\bibnamefont {Hayama}}, \bibinfo
  {author} {\bibfnamefont {V.}~\bibnamefont {Abdolazimi}}, \ and\ \bibinfo
  {author} {\bibfnamefont {H.}~\bibnamefont {Takagi}},\ }\href {\doibase
  10.1107/S2052520615006150} {\bibfield  {journal} {\bibinfo  {journal} {Acta
  Crystallographica Section B}\ }\textbf {\bibinfo {volume} {71}},\ \bibinfo
  {pages} {300} (\bibinfo {year} {2015})}\BibitemShut {NoStop}%
\bibitem [{\citenamefont {Dagotto}(2007)}]{when_oxides_meet}%
  \BibitemOpen
  \bibfield  {author} {\bibinfo {author} {\bibfnamefont {E.}~\bibnamefont
  {Dagotto}},\ }\href {\doibase 10.1126/science.1151094} {\bibfield  {journal}
  {\bibinfo  {journal} {Science}\ }\textbf {\bibinfo {volume} {318}},\ \bibinfo
  {pages} {1076} (\bibinfo {year} {2007})}\BibitemShut {NoStop}%
\bibitem [{\citenamefont {Zhu}\ \emph {et~al.}(2016)\citenamefont {Zhu},
  \citenamefont {Wang}, \citenamefont {Li}, \citenamefont {Howard},
  \citenamefont {Neuefeind}, \citenamefont {Ren}, \citenamefont {Wang},
  \citenamefont {Liang}, \citenamefont {Yang}, \citenamefont {Zou},
  \citenamefont {Jin},\ and\ \citenamefont {Zhao}}]{future_antiperovskites}%
  \BibitemOpen
  \bibfield  {author} {\bibinfo {author} {\bibfnamefont {J.}~\bibnamefont
  {Zhu}}, \bibinfo {author} {\bibfnamefont {Y.}~\bibnamefont {Wang}}, \bibinfo
  {author} {\bibfnamefont {S.}~\bibnamefont {Li}}, \bibinfo {author}
  {\bibfnamefont {J.~W.}\ \bibnamefont {Howard}}, \bibinfo {author}
  {\bibfnamefont {J.}~\bibnamefont {Neuefeind}}, \bibinfo {author}
  {\bibfnamefont {Y.}~\bibnamefont {Ren}}, \bibinfo {author} {\bibfnamefont
  {H.}~\bibnamefont {Wang}}, \bibinfo {author} {\bibfnamefont {C.}~\bibnamefont
  {Liang}}, \bibinfo {author} {\bibfnamefont {W.}~\bibnamefont {Yang}},
  \bibinfo {author} {\bibfnamefont {R.}~\bibnamefont {Zou}}, \bibinfo {author}
  {\bibfnamefont {C.}~\bibnamefont {Jin}}, \ and\ \bibinfo {author}
  {\bibfnamefont {Y.}~\bibnamefont {Zhao}},\ }\href {\doibase
  10.1021/acs.inorgchem.6b00444} {\bibfield  {journal} {\bibinfo  {journal}
  {Inorganic Chemistry}\ }\textbf {\bibinfo {volume} {55}},\ \bibinfo {pages}
  {5993} (\bibinfo {year} {2016})}\BibitemShut {NoStop}%
\bibitem [{\citenamefont {Kamishima}\ \emph {et~al.}(2000)\citenamefont
  {Kamishima}, \citenamefont {Goto}, \citenamefont {Nakagawa}, \citenamefont
  {Miura}, \citenamefont {Ohashi}, \citenamefont {Mori}, \citenamefont
  {Sasaki},\ and\ \citenamefont {Kanomata}}]{Magnetism_ant-perovskites}%
  \BibitemOpen
  \bibfield  {author} {\bibinfo {author} {\bibfnamefont {K.}~\bibnamefont
  {Kamishima}}, \bibinfo {author} {\bibfnamefont {T.}~\bibnamefont {Goto}},
  \bibinfo {author} {\bibfnamefont {H.}~\bibnamefont {Nakagawa}}, \bibinfo
  {author} {\bibfnamefont {N.}~\bibnamefont {Miura}}, \bibinfo {author}
  {\bibfnamefont {M.}~\bibnamefont {Ohashi}}, \bibinfo {author} {\bibfnamefont
  {N.}~\bibnamefont {Mori}}, \bibinfo {author} {\bibfnamefont {T.}~\bibnamefont
  {Sasaki}}, \ and\ \bibinfo {author} {\bibfnamefont {T.}~\bibnamefont
  {Kanomata}},\ }\href {\doibase 10.1103/PhysRevB.63.024426} {\bibfield
  {journal} {\bibinfo  {journal} {Phys. Rev. B}\ }\textbf {\bibinfo {volume}
  {63}},\ \bibinfo {pages} {024426} (\bibinfo {year} {2000})}\BibitemShut
  {NoStop}%
\bibitem [{\citenamefont {Wang}\ \emph {et~al.}(2009)\citenamefont {Wang},
  \citenamefont {Tong}, \citenamefont {Sun}, \citenamefont {Li}, \citenamefont
  {Tang}, \citenamefont {Lu}, \citenamefont {Zhu}, \citenamefont {Yang},\ and\
  \citenamefont {Song}}]{Magnetsm_anti1}%
  \BibitemOpen
  \bibfield  {author} {\bibinfo {author} {\bibfnamefont {B.~S.}\ \bibnamefont
  {Wang}}, \bibinfo {author} {\bibfnamefont {P.}~\bibnamefont {Tong}}, \bibinfo
  {author} {\bibfnamefont {Y.~P.}\ \bibnamefont {Sun}}, \bibinfo {author}
  {\bibfnamefont {L.~J.}\ \bibnamefont {Li}}, \bibinfo {author} {\bibfnamefont
  {W.}~\bibnamefont {Tang}}, \bibinfo {author} {\bibfnamefont {W.~J.}\
  \bibnamefont {Lu}}, \bibinfo {author} {\bibfnamefont {X.~B.}\ \bibnamefont
  {Zhu}}, \bibinfo {author} {\bibfnamefont {Z.~R.}\ \bibnamefont {Yang}}, \
  and\ \bibinfo {author} {\bibfnamefont {W.~H.}\ \bibnamefont {Song}},\ }\href
  {\doibase 10.1063/1.3268786} {\bibfield  {journal} {\bibinfo  {journal}
  {Applied Physics Letters}\ }\textbf {\bibinfo {volume} {95}},\ \bibinfo
  {pages} {222509} (\bibinfo {year} {2009})}\BibitemShut {NoStop}%
\bibitem [{\citenamefont {Iikubo}\ \emph {et~al.}(2008)\citenamefont {Iikubo},
  \citenamefont {Kodama}, \citenamefont {Takenaka}, \citenamefont {Takagi},
  \citenamefont {Takigawa},\ and\ \citenamefont {Shamoto}}]{Magnetsm_anti2}%
  \BibitemOpen
  \bibfield  {author} {\bibinfo {author} {\bibfnamefont {S.}~\bibnamefont
  {Iikubo}}, \bibinfo {author} {\bibfnamefont {K.}~\bibnamefont {Kodama}},
  \bibinfo {author} {\bibfnamefont {K.}~\bibnamefont {Takenaka}}, \bibinfo
  {author} {\bibfnamefont {H.}~\bibnamefont {Takagi}}, \bibinfo {author}
  {\bibfnamefont {M.}~\bibnamefont {Takigawa}}, \ and\ \bibinfo {author}
  {\bibfnamefont {S.}~\bibnamefont {Shamoto}},\ }\href {\doibase
  10.1103/PhysRevLett.101.205901} {\bibfield  {journal} {\bibinfo  {journal}
  {Phys. Rev. Lett.}\ }\textbf {\bibinfo {volume} {101}},\ \bibinfo {pages}
  {205901} (\bibinfo {year} {2008})}\BibitemShut {NoStop}%
\bibitem [{\citenamefont {Nakamura}\ \emph {et~al.}(2009)\citenamefont
  {Nakamura}, \citenamefont {Takenaka}, \citenamefont {Kishimoto},\ and\
  \citenamefont {Takagi}}]{electrical_thermal_conductivities}%
  \BibitemOpen
  \bibfield  {author} {\bibinfo {author} {\bibfnamefont {Y.}~\bibnamefont
  {Nakamura}}, \bibinfo {author} {\bibfnamefont {K.}~\bibnamefont {Takenaka}},
  \bibinfo {author} {\bibfnamefont {A.}~\bibnamefont {Kishimoto}}, \ and\
  \bibinfo {author} {\bibfnamefont {H.}~\bibnamefont {Takagi}},\ }\href
  {\doibase https://doi.org/10.1111/j.1551-2916.2009.03297.x} {\bibfield
  {journal} {\bibinfo  {journal} {Journal of the American Ceramic Society}\
  }\textbf {\bibinfo {volume} {92}},\ \bibinfo {pages} {2999} (\bibinfo {year}
  {2009})}\BibitemShut {NoStop}%
\bibitem [{\citenamefont {Hui}\ \emph {et~al.}(2014)\citenamefont {Hui},
  \citenamefont {Tang}, \citenamefont {Shao}, \citenamefont {Lei},
  \citenamefont {Yang}, \citenamefont {Song}, \citenamefont {Luo},
  \citenamefont {Zhu},\ and\ \citenamefont {Sun}}]{superconductivity_1}%
  \BibitemOpen
  \bibfield  {author} {\bibinfo {author} {\bibfnamefont {Z.}~\bibnamefont
  {Hui}}, \bibinfo {author} {\bibfnamefont {X.}~\bibnamefont {Tang}}, \bibinfo
  {author} {\bibfnamefont {D.}~\bibnamefont {Shao}}, \bibinfo {author}
  {\bibfnamefont {H.}~\bibnamefont {Lei}}, \bibinfo {author} {\bibfnamefont
  {J.}~\bibnamefont {Yang}}, \bibinfo {author} {\bibfnamefont {W.}~\bibnamefont
  {Song}}, \bibinfo {author} {\bibfnamefont {H.}~\bibnamefont {Luo}}, \bibinfo
  {author} {\bibfnamefont {X.}~\bibnamefont {Zhu}}, \ and\ \bibinfo {author}
  {\bibfnamefont {Y.}~\bibnamefont {Sun}},\ }\href {\doibase
  10.1039/C4CC05281A} {\bibfield  {journal} {\bibinfo  {journal} {Chem.
  Commun.}\ }\textbf {\bibinfo {volume} {50}},\ \bibinfo {pages} {12734}
  (\bibinfo {year} {2014})}\BibitemShut {NoStop}%
\bibitem [{\citenamefont {Lin}\ \emph {et~al.}(2015)\citenamefont {Lin},
  \citenamefont {Tong}, \citenamefont {Tong}, \citenamefont {Lin},
  \citenamefont {Wang}, \citenamefont {Song}, \citenamefont {Zou},\ and\
  \citenamefont {Sun}}]{thermal_expansion}%
  \BibitemOpen
  \bibfield  {author} {\bibinfo {author} {\bibfnamefont {J.~C.}\ \bibnamefont
  {Lin}}, \bibinfo {author} {\bibfnamefont {P.}~\bibnamefont {Tong}}, \bibinfo
  {author} {\bibfnamefont {W.}~\bibnamefont {Tong}}, \bibinfo {author}
  {\bibfnamefont {S.}~\bibnamefont {Lin}}, \bibinfo {author} {\bibfnamefont
  {B.~S.}\ \bibnamefont {Wang}}, \bibinfo {author} {\bibfnamefont {W.~H.}\
  \bibnamefont {Song}}, \bibinfo {author} {\bibfnamefont {Y.~M.}\ \bibnamefont
  {Zou}}, \ and\ \bibinfo {author} {\bibfnamefont {Y.~P.}\ \bibnamefont
  {Sun}},\ }\href {\doibase 10.1063/1.4913663} {\bibfield  {journal} {\bibinfo
  {journal} {Applied Physics Letters}\ }\textbf {\bibinfo {volume} {106}},\
  \bibinfo {pages} {082405} (\bibinfo {year} {2015})}\BibitemShut {NoStop}%
\bibitem [{\citenamefont {Vaughn~II}\ \emph {et~al.}(2014)\citenamefont
  {Vaughn~II}, \citenamefont {Araujo}, \citenamefont {Meduri}, \citenamefont
  {Callejas}, \citenamefont {Hickner},\ and\ \citenamefont
  {Schaak}}]{energy_conversion}%
  \BibitemOpen
  \bibfield  {author} {\bibinfo {author} {\bibfnamefont {D.~D.}\ \bibnamefont
  {Vaughn~II}}, \bibinfo {author} {\bibfnamefont {J.}~\bibnamefont {Araujo}},
  \bibinfo {author} {\bibfnamefont {P.}~\bibnamefont {Meduri}}, \bibinfo
  {author} {\bibfnamefont {J.~F.}\ \bibnamefont {Callejas}}, \bibinfo {author}
  {\bibfnamefont {M.~A.}\ \bibnamefont {Hickner}}, \ and\ \bibinfo {author}
  {\bibfnamefont {R.~E.}\ \bibnamefont {Schaak}},\ }\href {\doibase
  10.1021/cm5029723} {\bibfield  {journal} {\bibinfo  {journal} {Chemistry of
  Materials}\ }\textbf {\bibinfo {volume} {26}},\ \bibinfo {pages} {6226}
  (\bibinfo {year} {2014})}\BibitemShut {NoStop}%
\bibitem [{\citenamefont {Takenaka}\ \emph {et~al.}(2013)\citenamefont
  {Takenaka}, \citenamefont {Hamada}, \citenamefont {Shibayama},\ and\
  \citenamefont {Asano}}]{TAKENAKA2013S291}%
  \BibitemOpen
  \bibfield  {author} {\bibinfo {author} {\bibfnamefont {K.}~\bibnamefont
  {Takenaka}}, \bibinfo {author} {\bibfnamefont {T.}~\bibnamefont {Hamada}},
  \bibinfo {author} {\bibfnamefont {T.}~\bibnamefont {Shibayama}}, \ and\
  \bibinfo {author} {\bibfnamefont {K.}~\bibnamefont {Asano}},\ }\href
  {\doibase https://doi.org/10.1016/j.jallcom.2011.10.084} {\bibfield
  {journal} {\bibinfo  {journal} {Journal of Alloys and Compounds}\ }\textbf
  {\bibinfo {volume} {577}},\ \bibinfo {pages} {S291 } (\bibinfo {year}
  {2013})}\BibitemShut {NoStop}%
\bibitem [{\citenamefont {Fang}\ and\ \citenamefont
  {Cano}(2020)}]{High-order-topology-Anti-perovskites}%
  \BibitemOpen
  \bibfield  {author} {\bibinfo {author} {\bibfnamefont {Y.}~\bibnamefont
  {Fang}}\ and\ \bibinfo {author} {\bibfnamefont {J.}~\bibnamefont {Cano}},\
  }\href {\doibase 10.1103/PhysRevB.101.245110} {\bibfield  {journal} {\bibinfo
   {journal} {Phys. Rev. B}\ }\textbf {\bibinfo {volume} {101}},\ \bibinfo
  {pages} {245110} (\bibinfo {year} {2020})}\BibitemShut {NoStop}%
\bibitem [{\citenamefont {Zhao}\ and\ \citenamefont
  {Daemen}(2012)}]{energy_storage}%
  \BibitemOpen
  \bibfield  {author} {\bibinfo {author} {\bibfnamefont {Y.}~\bibnamefont
  {Zhao}}\ and\ \bibinfo {author} {\bibfnamefont {L.~L.}\ \bibnamefont
  {Daemen}},\ }\href {\doibase 10.1021/ja305709z} {\bibfield  {journal}
  {\bibinfo  {journal} {Journal of the American Chemical Society}\ }\textbf
  {\bibinfo {volume} {134}},\ \bibinfo {pages} {15042} (\bibinfo {year}
  {2012})}\BibitemShut {NoStop}%
\bibitem [{\citenamefont {Evans}\ \emph {et~al.}(2020)\citenamefont {Evans},
  \citenamefont {Wu}, \citenamefont {Seshadri},\ and\ \citenamefont
  {Cheetham}}]{Evans2020_antiperovskite}%
  \BibitemOpen
  \bibfield  {author} {\bibinfo {author} {\bibfnamefont {H.~A.}\ \bibnamefont
  {Evans}}, \bibinfo {author} {\bibfnamefont {Y.}~\bibnamefont {Wu}}, \bibinfo
  {author} {\bibfnamefont {R.}~\bibnamefont {Seshadri}}, \ and\ \bibinfo
  {author} {\bibfnamefont {A.~K.}\ \bibnamefont {Cheetham}},\ }\href {\doibase
  10.1038/s41578-019-0160-x} {\bibfield  {journal} {\bibinfo  {journal} {Nature
  Reviews Materials}\ }\textbf {\bibinfo {volume} {5}},\ \bibinfo {pages} {196}
  (\bibinfo {year} {2020})}\BibitemShut {NoStop}%
\bibitem [{\citenamefont {Ullah}\ \emph {et~al.}(2016)\citenamefont {Ullah},
  \citenamefont {Murtaza}, \citenamefont {Khenata}, \citenamefont {Mahmood},
  \citenamefont {Muzzamil}, \citenamefont {Amin},\ and\ \citenamefont
  {Saleh}}]{Ullah_2016}%
  \BibitemOpen
  \bibfield  {author} {\bibinfo {author} {\bibfnamefont {I.}~\bibnamefont
  {Ullah}}, \bibinfo {author} {\bibfnamefont {G.}~\bibnamefont {Murtaza}},
  \bibinfo {author} {\bibfnamefont {R.}~\bibnamefont {Khenata}}, \bibinfo
  {author} {\bibfnamefont {A.}~\bibnamefont {Mahmood}}, \bibinfo {author}
  {\bibfnamefont {M.}~\bibnamefont {Muzzamil}}, \bibinfo {author}
  {\bibfnamefont {N.}~\bibnamefont {Amin}}, \ and\ \bibinfo {author}
  {\bibfnamefont {M.}~\bibnamefont {Saleh}},\ }\href {\doibase
  10.1007/s11664-015-4330-3} {\bibfield  {journal} {\bibinfo  {journal}
  {Journal of Electronic Materials}\ }\textbf {\bibinfo {volume} {45}},\
  \bibinfo {pages} {3059} (\bibinfo {year} {2016})}\BibitemShut {NoStop}%
\bibitem [{\citenamefont {Ochi}\ and\ \citenamefont
  {Kuroki}(2019{\natexlab{b}})}]{hexagonal_anti-perov}%
  \BibitemOpen
  \bibfield  {author} {\bibinfo {author} {\bibfnamefont {M.}~\bibnamefont
  {Ochi}}\ and\ \bibinfo {author} {\bibfnamefont {K.}~\bibnamefont {Kuroki}},\
  }\href {\doibase 10.1103/PhysRevApplied.12.034009} {\bibfield  {journal}
  {\bibinfo  {journal} {Phys. Rev. Applied}\ }\textbf {\bibinfo {volume}
  {12}},\ \bibinfo {pages} {034009} (\bibinfo {year}
  {2019}{\natexlab{b}})}\BibitemShut {NoStop}%
\bibitem [{\citenamefont {Mochizuki}\ \emph {et~al.}(2020)\citenamefont
  {Mochizuki}, \citenamefont {Sung}, \citenamefont {Takahashi}, \citenamefont
  {Kumagai},\ and\ \citenamefont {Oba}}]{semiconducting_inverse-perovs}%
  \BibitemOpen
  \bibfield  {author} {\bibinfo {author} {\bibfnamefont {Y.}~\bibnamefont
  {Mochizuki}}, \bibinfo {author} {\bibfnamefont {H.-J.}\ \bibnamefont {Sung}},
  \bibinfo {author} {\bibfnamefont {A.}~\bibnamefont {Takahashi}}, \bibinfo
  {author} {\bibfnamefont {Y.}~\bibnamefont {Kumagai}}, \ and\ \bibinfo
  {author} {\bibfnamefont {F.}~\bibnamefont {Oba}},\ }\href {\doibase
  10.1103/PhysRevMaterials.4.044601} {\bibfield  {journal} {\bibinfo  {journal}
  {Phys. Rev. Materials}\ }\textbf {\bibinfo {volume} {4}},\ \bibinfo {pages}
  {044601} (\bibinfo {year} {2020})}\BibitemShut {NoStop}%
\bibitem [{\citenamefont {Soler}\ \emph {et~al.}(2002)\citenamefont {Soler},
  \citenamefont {Artacho}, \citenamefont {Gale}, \citenamefont {Garc{\'{\i}}a},
  \citenamefont {Junquera}, \citenamefont {Ordej{\'{o}}n},\ and\ \citenamefont
  {S{\'{a}}nchez-Portal}}]{SIESTA}%
  \BibitemOpen
  \bibfield  {author} {\bibinfo {author} {\bibfnamefont {J.~M.}\ \bibnamefont
  {Soler}}, \bibinfo {author} {\bibfnamefont {E.}~\bibnamefont {Artacho}},
  \bibinfo {author} {\bibfnamefont {J.~D.}\ \bibnamefont {Gale}}, \bibinfo
  {author} {\bibfnamefont {A.}~\bibnamefont {Garc{\'{\i}}a}}, \bibinfo {author}
  {\bibfnamefont {J.}~\bibnamefont {Junquera}}, \bibinfo {author}
  {\bibfnamefont {P.}~\bibnamefont {Ordej{\'{o}}n}}, \ and\ \bibinfo {author}
  {\bibfnamefont {D.}~\bibnamefont {S{\'{a}}nchez-Portal}},\ }\href {\doibase
  10.1088/0953-8984/14/11/302} {\bibfield  {journal} {\bibinfo  {journal} {J.
  Phys.: Condens. Matter}\ }\textbf {\bibinfo {volume} {14}},\ \bibinfo {pages}
  {2745} (\bibinfo {year} {2002})}\BibitemShut {NoStop}%
\bibitem [{\citenamefont {Bao}\ \emph {et~al.}(2018)\citenamefont {Bao},
  \citenamefont {Gagliardi},\ and\ \citenamefont {Truhlar}}]{sic}%
  \BibitemOpen
  \bibfield  {author} {\bibinfo {author} {\bibfnamefont {J.~L.}\ \bibnamefont
  {Bao}}, \bibinfo {author} {\bibfnamefont {L.}~\bibnamefont {Gagliardi}}, \
  and\ \bibinfo {author} {\bibfnamefont {D.~G.}\ \bibnamefont {Truhlar}},\
  }\href {\doibase 10.1021/acs.jpclett.8b00242} {\bibfield  {journal} {\bibinfo
   {journal} {The Journal of Physical Chemistry Letters}\ }\textbf {\bibinfo
  {volume} {9}},\ \bibinfo {pages} {2353} (\bibinfo {year} {2018})},\ \bibinfo
  {note} {pMID: 29624392}\BibitemShut {NoStop}%
\bibitem [{\citenamefont {Mosquera}\ \emph {et~al.}(2016)\citenamefont
  {Mosquera}, \citenamefont {Borca}, \citenamefont {Ratner},\ and\
  \citenamefont {Schatz}}]{Hybrids}%
  \BibitemOpen
  \bibfield  {author} {\bibinfo {author} {\bibfnamefont {M.~A.}\ \bibnamefont
  {Mosquera}}, \bibinfo {author} {\bibfnamefont {C.~H.}\ \bibnamefont {Borca}},
  \bibinfo {author} {\bibfnamefont {M.~A.}\ \bibnamefont {Ratner}}, \ and\
  \bibinfo {author} {\bibfnamefont {G.~C.}\ \bibnamefont {Schatz}},\ }\href
  {\doibase 10.1021/acs.jpca.5b10864} {\bibfield  {journal} {\bibinfo
  {journal} {The Journal of Physical Chemistry A}\ }\textbf {\bibinfo {volume}
  {120}},\ \bibinfo {pages} {1605} (\bibinfo {year} {2016})}\BibitemShut
  {NoStop}%
\bibitem [{\citenamefont {Anisimov}\ \emph {et~al.}(1991)\citenamefont
  {Anisimov}, \citenamefont {Zaanen},\ and\ \citenamefont {Andersen}}]{LDA+U}%
  \BibitemOpen
  \bibfield  {author} {\bibinfo {author} {\bibfnamefont {V.~I.}\ \bibnamefont
  {Anisimov}}, \bibinfo {author} {\bibfnamefont {J.}~\bibnamefont {Zaanen}}, \
  and\ \bibinfo {author} {\bibfnamefont {O.~K.}\ \bibnamefont {Andersen}},\
  }\href {\doibase 10.1103/PhysRevB.44.943} {\bibfield  {journal} {\bibinfo
  {journal} {Phys. Rev. B}\ }\textbf {\bibinfo {volume} {44}},\ \bibinfo
  {pages} {943} (\bibinfo {year} {1991})}\BibitemShut {NoStop}%
\bibitem [{\citenamefont {Perdew}\ \emph {et~al.}(1996)\citenamefont {Perdew},
  \citenamefont {Burke},\ and\ \citenamefont {Ernzerhof}}]{GGA-1996}%
  \BibitemOpen
  \bibfield  {author} {\bibinfo {author} {\bibfnamefont {J.~P.}\ \bibnamefont
  {Perdew}}, \bibinfo {author} {\bibfnamefont {K.}~\bibnamefont {Burke}}, \
  and\ \bibinfo {author} {\bibfnamefont {M.}~\bibnamefont {Ernzerhof}},\ }\href
  {\doibase 10.1103/PhysRevLett.77.3865} {\bibfield  {journal} {\bibinfo
  {journal} {Phys. Rev. Lett.}\ }\textbf {\bibinfo {volume} {77}},\ \bibinfo
  {pages} {3865} (\bibinfo {year} {1996})}\BibitemShut {NoStop}%
\bibitem [{\citenamefont {Hohenberg}\ and\ \citenamefont
  {Kohn}(1964)}]{Hohenberg-64}%
  \BibitemOpen
  \bibfield  {author} {\bibinfo {author} {\bibfnamefont {P.}~\bibnamefont
  {Hohenberg}}\ and\ \bibinfo {author} {\bibfnamefont {W.}~\bibnamefont
  {Kohn}},\ }\href {\doibase 10.1103/PhysRev.136.B864} {\bibfield  {journal}
  {\bibinfo  {journal} {Phys. Rev.}\ }\textbf {\bibinfo {volume} {136}},\
  \bibinfo {pages} {B864} (\bibinfo {year} {1964})}\BibitemShut {NoStop}%
\bibitem [{\citenamefont {Kohn}\ and\ \citenamefont {Sham}(1965)}]{Kohn-65}%
  \BibitemOpen
  \bibfield  {author} {\bibinfo {author} {\bibfnamefont {W.}~\bibnamefont
  {Kohn}}\ and\ \bibinfo {author} {\bibfnamefont {L.~J.}\ \bibnamefont
  {Sham}},\ }\href {\doibase 10.1103/PhysRev.140.A1133} {\bibfield  {journal}
  {\bibinfo  {journal} {Phys. Rev.}\ }\textbf {\bibinfo {volume} {140}},\
  \bibinfo {pages} {A1133} (\bibinfo {year} {1965})}\BibitemShut {NoStop}%
\bibitem [{\citenamefont {Johnson}\ and\ \citenamefont
  {Ashcroft}(1998)}]{DFT_Bandgap_problem}%
  \BibitemOpen
  \bibfield  {author} {\bibinfo {author} {\bibfnamefont {K.~A.}\ \bibnamefont
  {Johnson}}\ and\ \bibinfo {author} {\bibfnamefont {N.~W.}\ \bibnamefont
  {Ashcroft}},\ }\href {\doibase 10.1103/PhysRevB.58.15548} {\bibfield
  {journal} {\bibinfo  {journal} {Phys. Rev. B}\ }\textbf {\bibinfo {volume}
  {58}},\ \bibinfo {pages} {15548} (\bibinfo {year} {1998})}\BibitemShut
  {NoStop}%
\bibitem [{\citenamefont {Krukau}\ \emph {et~al.}(2006)\citenamefont {Krukau},
  \citenamefont {Vydrov}, \citenamefont {Izmaylov},\ and\ \citenamefont
  {Scuseria}}]{HSE06}%
  \BibitemOpen
  \bibfield  {author} {\bibinfo {author} {\bibfnamefont {A.~V.}\ \bibnamefont
  {Krukau}}, \bibinfo {author} {\bibfnamefont {O.~A.}\ \bibnamefont {Vydrov}},
  \bibinfo {author} {\bibfnamefont {A.~F.}\ \bibnamefont {Izmaylov}}, \ and\
  \bibinfo {author} {\bibfnamefont {G.~E.}\ \bibnamefont {Scuseria}},\ }\href
  {\doibase 10.1063/1.2404663} {\bibfield  {journal} {\bibinfo  {journal} {The
  Journal of Chemical Physics}\ }\textbf {\bibinfo {volume} {125}},\ \bibinfo
  {pages} {224106} (\bibinfo {year} {2006})}\BibitemShut {NoStop}%
\bibitem [{\citenamefont {Garc\'{\i}a}\ \emph {et~al.}(2020)\citenamefont
  {Garc\'{\i}a}, \citenamefont {Papior}, \citenamefont {Akhtar}, \citenamefont
  {Artacho}, \citenamefont {Blum}, \citenamefont {Bosoni}, \citenamefont
  {Brandimarte}, \citenamefont {Brandbyge}, \citenamefont {Cerd\'a},
  \citenamefont {Corsetti}, \citenamefont {Cuadrado}, \citenamefont {Dikan},
  \citenamefont {Ferrer}, \citenamefont {Gale}, \citenamefont
  {Garc\'{\i}a-Fern\'andez}, \citenamefont {Garc\'{\i}a-Su\'arez},
  \citenamefont {Garc\'{\i}a}, \citenamefont {Huhs}, \citenamefont {Illera},
  \citenamefont {Koryt\'ar}, \citenamefont {Koval}, \citenamefont {Lebedeva},
  \citenamefont {Lin}, \citenamefont {L\'opez-Tarifa}, \citenamefont {Mayo},
  \citenamefont {Mohr}, \citenamefont {Ordej\'on}, \citenamefont {Postnikov},
  \citenamefont {Pouillon}, \citenamefont {Pruneda}, \citenamefont {Robles},
  \citenamefont {S\'anchez-Portal}, \citenamefont {Soler}, \citenamefont
  {Ullah}, \citenamefont {Yu},\ and\ \citenamefont {Junquera}}]{New_SIESTA}%
  \BibitemOpen
  \bibfield  {author} {\bibinfo {author} {\bibfnamefont {A.}~\bibnamefont
  {Garc\'{\i}a}}, \bibinfo {author} {\bibfnamefont {N.}~\bibnamefont {Papior}},
  \bibinfo {author} {\bibfnamefont {A.}~\bibnamefont {Akhtar}}, \bibinfo
  {author} {\bibfnamefont {E.}~\bibnamefont {Artacho}}, \bibinfo {author}
  {\bibfnamefont {V.}~\bibnamefont {Blum}}, \bibinfo {author} {\bibfnamefont
  {E.}~\bibnamefont {Bosoni}}, \bibinfo {author} {\bibfnamefont
  {P.}~\bibnamefont {Brandimarte}}, \bibinfo {author} {\bibfnamefont
  {M.}~\bibnamefont {Brandbyge}}, \bibinfo {author} {\bibfnamefont {J.~I.}\
  \bibnamefont {Cerd\'a}}, \bibinfo {author} {\bibfnamefont {F.}~\bibnamefont
  {Corsetti}}, \bibinfo {author} {\bibfnamefont {R.}~\bibnamefont {Cuadrado}},
  \bibinfo {author} {\bibfnamefont {V.}~\bibnamefont {Dikan}}, \bibinfo
  {author} {\bibfnamefont {J.}~\bibnamefont {Ferrer}}, \bibinfo {author}
  {\bibfnamefont {J.}~\bibnamefont {Gale}}, \bibinfo {author} {\bibfnamefont
  {P.}~\bibnamefont {Garc\'{\i}a-Fern\'andez}}, \bibinfo {author}
  {\bibfnamefont {V.~M.}\ \bibnamefont {Garc\'{\i}a-Su\'arez}}, \bibinfo
  {author} {\bibfnamefont {S.}~\bibnamefont {Garc\'{\i}a}}, \bibinfo {author}
  {\bibfnamefont {G.}~\bibnamefont {Huhs}}, \bibinfo {author} {\bibfnamefont
  {S.}~\bibnamefont {Illera}}, \bibinfo {author} {\bibfnamefont
  {R.}~\bibnamefont {Koryt\'ar}}, \bibinfo {author} {\bibfnamefont
  {P.}~\bibnamefont {Koval}}, \bibinfo {author} {\bibfnamefont
  {I.}~\bibnamefont {Lebedeva}}, \bibinfo {author} {\bibfnamefont
  {L.}~\bibnamefont {Lin}}, \bibinfo {author} {\bibfnamefont {P.}~\bibnamefont
  {L\'opez-Tarifa}}, \bibinfo {author} {\bibfnamefont {S.~G.}\ \bibnamefont
  {Mayo}}, \bibinfo {author} {\bibfnamefont {S.}~\bibnamefont {Mohr}}, \bibinfo
  {author} {\bibfnamefont {P.}~\bibnamefont {Ordej\'on}}, \bibinfo {author}
  {\bibfnamefont {A.}~\bibnamefont {Postnikov}}, \bibinfo {author}
  {\bibfnamefont {Y.}~\bibnamefont {Pouillon}}, \bibinfo {author}
  {\bibfnamefont {M.}~\bibnamefont {Pruneda}}, \bibinfo {author} {\bibfnamefont
  {R.}~\bibnamefont {Robles}}, \bibinfo {author} {\bibfnamefont
  {D.}~\bibnamefont {S\'anchez-Portal}}, \bibinfo {author} {\bibfnamefont
  {J.~M.}\ \bibnamefont {Soler}}, \bibinfo {author} {\bibfnamefont
  {R.}~\bibnamefont {Ullah}}, \bibinfo {author} {\bibfnamefont {V.~W.~z.}\
  \bibnamefont {Yu}}, \ and\ \bibinfo {author} {\bibfnamefont {J.}~\bibnamefont
  {Junquera}},\ }\href {\doibase 10.1063/5.0005077} {\bibfield  {journal}
  {\bibinfo  {journal} {The Journal of Chemical Physics}\ }\textbf {\bibinfo
  {volume} {152}},\ \bibinfo {pages} {204108} (\bibinfo {year}
  {2020})}\BibitemShut {NoStop}%
\bibitem [{\citenamefont {Troullier}\ and\ \citenamefont
  {Martins}(1991)}]{Troullier-91}%
  \BibitemOpen
  \bibfield  {author} {\bibinfo {author} {\bibfnamefont {N.}~\bibnamefont
  {Troullier}}\ and\ \bibinfo {author} {\bibfnamefont {J.~L.}\ \bibnamefont
  {Martins}},\ }\href {\doibase 10.1103/PhysRevB.43.1993} {\bibfield  {journal}
  {\bibinfo  {journal} {Phys. Rev. B}\ }\textbf {\bibinfo {volume} {43}},\
  \bibinfo {pages} {1993} (\bibinfo {year} {1991})}\BibitemShut {NoStop}%
\bibitem [{\citenamefont {Kleinman}\ and\ \citenamefont
  {Bylander}(1982)}]{Kleinman-82}%
  \BibitemOpen
  \bibfield  {author} {\bibinfo {author} {\bibfnamefont {L.}~\bibnamefont
  {Kleinman}}\ and\ \bibinfo {author} {\bibfnamefont {D.~M.}\ \bibnamefont
  {Bylander}},\ }\href {\doibase 10.1103/PhysRevLett.48.1425} {\bibfield
  {journal} {\bibinfo  {journal} {Phys. Rev. Lett.}\ }\textbf {\bibinfo
  {volume} {48}},\ \bibinfo {pages} {1425} (\bibinfo {year}
  {1982})}\BibitemShut {NoStop}%
\bibitem [{\citenamefont {Junquera}\ \emph {et~al.}(2003)\citenamefont
  {Junquera}, \citenamefont {Zimmer}, \citenamefont {Ordej\'on},\ and\
  \citenamefont {Ghosez}}]{Junquera-03.2}%
  \BibitemOpen
  \bibfield  {author} {\bibinfo {author} {\bibfnamefont {J.}~\bibnamefont
  {Junquera}}, \bibinfo {author} {\bibfnamefont {M.}~\bibnamefont {Zimmer}},
  \bibinfo {author} {\bibfnamefont {P.}~\bibnamefont {Ordej\'on}}, \ and\
  \bibinfo {author} {\bibfnamefont {P.}~\bibnamefont {Ghosez}},\ }\href
  {\doibase 10.1103/PhysRevB.67.155327} {\bibfield  {journal} {\bibinfo
  {journal} {Phys. Rev. B}\ }\textbf {\bibinfo {volume} {67}},\ \bibinfo
  {pages} {155327} (\bibinfo {year} {2003})}\BibitemShut {NoStop}%
\bibitem [{\citenamefont {Chege}\ \emph {et~al.}(2020)\citenamefont {Chege},
  \citenamefont {Ning’i}, \citenamefont {Sifuna},\ and\ \citenamefont
  {Amolo}}]{Chege}%
  \BibitemOpen
  \bibfield  {author} {\bibinfo {author} {\bibfnamefont {S.}~\bibnamefont
  {Chege}}, \bibinfo {author} {\bibfnamefont {P.}~\bibnamefont {Ning’i}},
  \bibinfo {author} {\bibfnamefont {J.}~\bibnamefont {Sifuna}}, \ and\ \bibinfo
  {author} {\bibfnamefont {G.~O.}\ \bibnamefont {Amolo}},\ }\href {\doibase
  10.1063/5.0022525} {\bibfield  {journal} {\bibinfo  {journal} {AIP Advances}\
  }\textbf {\bibinfo {volume} {10}},\ \bibinfo {pages} {095018} (\bibinfo
  {year} {2020})}\BibitemShut {NoStop}%
\bibitem [{\citenamefont {Ning'i}\ \emph {et~al.}(2020)\citenamefont {Ning'i},
  \citenamefont {Chege}, \citenamefont {Sifuna},\ and\ \citenamefont
  {Amolo}}]{ningi2020interplay}%
  \BibitemOpen
  \bibfield  {author} {\bibinfo {author} {\bibfnamefont {P.}~\bibnamefont
  {Ning'i}}, \bibinfo {author} {\bibfnamefont {S.}~\bibnamefont {Chege}},
  \bibinfo {author} {\bibfnamefont {J.}~\bibnamefont {Sifuna}}, \ and\ \bibinfo
  {author} {\bibfnamefont {G.}~\bibnamefont {Amolo}},\ }\href@noop {} {\
  (\bibinfo {year} {2020})},\ \Eprint {http://arxiv.org/abs/2009.02529}
  {arXiv:2009.02529 [cond-mat.mtrl-sci]} \BibitemShut {NoStop}%
\bibitem [{\citenamefont {Sifuna}\ \emph
  {et~al.}(2020{\natexlab{a}})\citenamefont {Sifuna}, \citenamefont {Manyali},
  \citenamefont {Wabululu}, \citenamefont {Songa}, \citenamefont {Otieno},\
  and\ \citenamefont {Sironik}}]{sifuna2020ab}%
  \BibitemOpen
  \bibfield  {author} {\bibinfo {author} {\bibfnamefont {J.}~\bibnamefont
  {Sifuna}}, \bibinfo {author} {\bibfnamefont {G.~S.}\ \bibnamefont {Manyali}},
  \bibinfo {author} {\bibfnamefont {E.}~\bibnamefont {Wabululu}}, \bibinfo
  {author} {\bibfnamefont {C.}~\bibnamefont {Songa}}, \bibinfo {author}
  {\bibfnamefont {A.}~\bibnamefont {Otieno}}, \ and\ \bibinfo {author}
  {\bibfnamefont {S.}~\bibnamefont {Sironik}},\ }\href@noop {} {\  (\bibinfo
  {year} {2020}{\natexlab{a}})},\ \Eprint {http://arxiv.org/abs/2001.01488}
  {arXiv:2001.01488 [physics.comp-ph]} \BibitemShut {NoStop}%
\bibitem [{\citenamefont {Sankey}\ and\ \citenamefont
  {Niklewski}(1989)}]{Sankey-89}%
  \BibitemOpen
  \bibfield  {author} {\bibinfo {author} {\bibfnamefont {O.~F.}\ \bibnamefont
  {Sankey}}\ and\ \bibinfo {author} {\bibfnamefont {D.~J.}\ \bibnamefont
  {Niklewski}},\ }\href {\doibase 10.1103/PhysRevB.40.3979} {\bibfield
  {journal} {\bibinfo  {journal} {Phys. Rev. B}\ }\textbf {\bibinfo {volume}
  {40}},\ \bibinfo {pages} {3979} (\bibinfo {year} {1989})}\BibitemShut
  {NoStop}%
\bibitem [{\citenamefont {Artacho}\ \emph {et~al.}(1999)\citenamefont
  {Artacho}, \citenamefont {S{\'{a}}nchez-Portal}, \citenamefont
  {Ordej{\'{o}}n}, \citenamefont {Garc{\'{\i}}a},\ and\ \citenamefont
  {Soler}}]{Artacho-99}%
  \BibitemOpen
  \bibfield  {author} {\bibinfo {author} {\bibfnamefont {E.}~\bibnamefont
  {Artacho}}, \bibinfo {author} {\bibfnamefont {D.}~\bibnamefont
  {S{\'{a}}nchez-Portal}}, \bibinfo {author} {\bibfnamefont {P.}~\bibnamefont
  {Ordej{\'{o}}n}}, \bibinfo {author} {\bibfnamefont {A.}~\bibnamefont
  {Garc{\'{\i}}a}}, \ and\ \bibinfo {author} {\bibfnamefont {J.~M.}\
  \bibnamefont {Soler}},\ }\href {\doibase
  10.1002/(SICI)1521-3951(199909)215:1<809::AID-PSSB809>3.0.CO;2-0} {\bibfield
  {journal} {\bibinfo  {journal} {Phys. Status Solidi (b)}\ }\textbf {\bibinfo
  {volume} {215}},\ \bibinfo {pages} {809} (\bibinfo {year}
  {1999})}\BibitemShut {NoStop}%
\bibitem [{\citenamefont {Monkhorst}\ and\ \citenamefont
  {Pack}(1976)}]{Monkhorst-76}%
  \BibitemOpen
  \bibfield  {author} {\bibinfo {author} {\bibfnamefont {H.~J.}\ \bibnamefont
  {Monkhorst}}\ and\ \bibinfo {author} {\bibfnamefont {J.~D.}\ \bibnamefont
  {Pack}},\ }\href {\doibase 10.1103/PhysRevB.13.5188} {\bibfield  {journal}
  {\bibinfo  {journal} {Phys. Rev. B}\ }\textbf {\bibinfo {volume} {13}},\
  \bibinfo {pages} {5188} (\bibinfo {year} {1976})}\BibitemShut {NoStop}%
\bibitem [{\citenamefont {Tromer}\ \emph {et~al.}(2021)\citenamefont {Tromer},
  \citenamefont {Felix}, \citenamefont {Woellner},\ and\ \citenamefont
  {Galvao}}]{TROMER2021138210}%
  \BibitemOpen
  \bibfield  {author} {\bibinfo {author} {\bibfnamefont {R.~M.}\ \bibnamefont
  {Tromer}}, \bibinfo {author} {\bibfnamefont {L.~C.}\ \bibnamefont {Felix}},
  \bibinfo {author} {\bibfnamefont {C.~F.}\ \bibnamefont {Woellner}}, \ and\
  \bibinfo {author} {\bibfnamefont {D.~S.}\ \bibnamefont {Galvao}},\ }\href
  {\doibase https://doi.org/10.1016/j.cplett.2020.138210} {\bibfield  {journal}
  {\bibinfo  {journal} {Chemical Physics Letters}\ }\textbf {\bibinfo {volume}
  {763}},\ \bibinfo {pages} {138210} (\bibinfo {year} {2021})}\BibitemShut
  {NoStop}%
\bibitem [{\citenamefont {King-Smith}\ and\ \citenamefont
  {Vanderbilt}(1993)}]{king-smith-polarization}%
  \BibitemOpen
  \bibfield  {author} {\bibinfo {author} {\bibfnamefont {R.~D.}\ \bibnamefont
  {King-Smith}}\ and\ \bibinfo {author} {\bibfnamefont {D.}~\bibnamefont
  {Vanderbilt}},\ }\href {\doibase 10.1103/PhysRevB.47.1651} {\bibfield
  {journal} {\bibinfo  {journal} {Phys. Rev. B}\ }\textbf {\bibinfo {volume}
  {47}},\ \bibinfo {pages} {1651} (\bibinfo {year} {1993})}\BibitemShut
  {NoStop}%
\bibitem [{\citenamefont {Murnaghan}(1944)}]{Murnaghan244}%
  \BibitemOpen
  \bibfield  {author} {\bibinfo {author} {\bibfnamefont {F.~D.}\ \bibnamefont
  {Murnaghan}},\ }\href {\doibase 10.1073/pnas.30.9.244} {\bibfield  {journal}
  {\bibinfo  {journal} {Proceedings of the National Academy of Sciences}\
  }\textbf {\bibinfo {volume} {30}},\ \bibinfo {pages} {244} (\bibinfo {year}
  {1944})}\BibitemShut {NoStop}%
\bibitem [{\citenamefont {Manyali}\ and\ \citenamefont
  {Sifuna}(2019)}]{Low_comp}%
  \BibitemOpen
  \bibfield  {author} {\bibinfo {author} {\bibfnamefont {G.~S.}\ \bibnamefont
  {Manyali}}\ and\ \bibinfo {author} {\bibfnamefont {J.}~\bibnamefont
  {Sifuna}},\ }\href {\doibase 10.1063/1.5129268} {\bibfield  {journal}
  {\bibinfo  {journal} {AIP Advances}\ }\textbf {\bibinfo {volume} {9}},\
  \bibinfo {pages} {125029} (\bibinfo {year} {2019})}\BibitemShut {NoStop}%
\bibitem [{\citenamefont {Schlom}\ \emph {et~al.}(2007)\citenamefont {Schlom},
  \citenamefont {Chen}, \citenamefont {Eom}, \citenamefont {Rabe},
  \citenamefont {Streiffer},\ and\ \citenamefont
  {Triscone}}]{interfaces_oxides}%
  \BibitemOpen
  \bibfield  {author} {\bibinfo {author} {\bibfnamefont {D.~G.}\ \bibnamefont
  {Schlom}}, \bibinfo {author} {\bibfnamefont {L.-Q.}\ \bibnamefont {Chen}},
  \bibinfo {author} {\bibfnamefont {C.-B.}\ \bibnamefont {Eom}}, \bibinfo
  {author} {\bibfnamefont {K.~M.}\ \bibnamefont {Rabe}}, \bibinfo {author}
  {\bibfnamefont {S.~K.}\ \bibnamefont {Streiffer}}, \ and\ \bibinfo {author}
  {\bibfnamefont {J.-M.}\ \bibnamefont {Triscone}},\ }\href {\doibase
  10.1146/annurev.matsci.37.061206.113016} {\bibfield  {journal} {\bibinfo
  {journal} {Annual Review of Materials Research}\ }\textbf {\bibinfo {volume}
  {37}},\ \bibinfo {pages} {589} (\bibinfo {year} {2007})}\BibitemShut
  {NoStop}%
\bibitem [{\citenamefont {Iqbal}\ \emph {et~al.}(2016)\citenamefont {Iqbal},
  \citenamefont {Murtaza}, \citenamefont {Khenata}, \citenamefont {Mahmood},
  \citenamefont {Yar}, \citenamefont {Muzammil},\ and\ \citenamefont
  {Khan}}]{springer_nitrides}%
  \BibitemOpen
  \bibfield  {author} {\bibinfo {author} {\bibfnamefont {S.}~\bibnamefont
  {Iqbal}}, \bibinfo {author} {\bibfnamefont {G.}~\bibnamefont {Murtaza}},
  \bibinfo {author} {\bibfnamefont {R.}~\bibnamefont {Khenata}}, \bibinfo
  {author} {\bibfnamefont {A.}~\bibnamefont {Mahmood}}, \bibinfo {author}
  {\bibfnamefont {A.}~\bibnamefont {Yar}}, \bibinfo {author} {\bibfnamefont
  {M.}~\bibnamefont {Muzammil}}, \ and\ \bibinfo {author} {\bibfnamefont
  {M.}~\bibnamefont {Khan}},\ }\href {\doibase 10.1007/s11664-016-4563-9}
  {\bibfield  {journal} {\bibinfo  {journal} {Journal of Electronic Materials}\
  }\textbf {\bibinfo {volume} {45}},\ \bibinfo {pages} {4188} (\bibinfo {year}
  {2016})}\BibitemShut {NoStop}%
\bibitem [{\citenamefont {G{\"a}bler}\ \emph {et~al.}(2004)\citenamefont
  {G{\"a}bler}, \citenamefont {Kirchner}, \citenamefont {Schnelle},
  \citenamefont {Schwarz}, \citenamefont {Schmitt}, \citenamefont {Rosner},\
  and\ \citenamefont {Niewa}}]{Sr3Bi/SbN}%
  \BibitemOpen
  \bibfield  {author} {\bibinfo {author} {\bibfnamefont {F.}~\bibnamefont
  {G{\"a}bler}}, \bibinfo {author} {\bibfnamefont {M.}~\bibnamefont
  {Kirchner}}, \bibinfo {author} {\bibfnamefont {W.}~\bibnamefont {Schnelle}},
  \bibinfo {author} {\bibfnamefont {U.}~\bibnamefont {Schwarz}}, \bibinfo
  {author} {\bibfnamefont {M.}~\bibnamefont {Schmitt}}, \bibinfo {author}
  {\bibfnamefont {H.}~\bibnamefont {Rosner}}, \ and\ \bibinfo {author}
  {\bibfnamefont {R.}~\bibnamefont {Niewa}},\ }\href {\doibase
  10.1002/zaac.200400256} {\bibfield  {journal} {\bibinfo  {journal}
  {Zeitschrift für anorganische und allgemeine Chemie}\ }\textbf {\bibinfo
  {volume} {630}},\ \bibinfo {pages} {2292} (\bibinfo {year}
  {2004})}\BibitemShut {NoStop}%
\bibitem [{\citenamefont {Ghosez}\ \emph {et~al.}(1998)\citenamefont {Ghosez},
  \citenamefont {Michenaud},\ and\ \citenamefont {Gonze}}]{Born_effective}%
  \BibitemOpen
  \bibfield  {author} {\bibinfo {author} {\bibfnamefont {P.}~\bibnamefont
  {Ghosez}}, \bibinfo {author} {\bibfnamefont {J.-P.}\ \bibnamefont
  {Michenaud}}, \ and\ \bibinfo {author} {\bibfnamefont {X.}~\bibnamefont
  {Gonze}},\ }\href {\doibase 10.1103/PhysRevB.58.6224} {\bibfield  {journal}
  {\bibinfo  {journal} {Phys. Rev. B}\ }\textbf {\bibinfo {volume} {58}},\
  \bibinfo {pages} {6224} (\bibinfo {year} {1998})}\BibitemShut {NoStop}%
\bibitem [{\citenamefont {Ghosez}(1997)}]{Ghosez-thesis}%
  \BibitemOpen
  \bibfield  {author} {\bibinfo {author} {\bibfnamefont {P.}~\bibnamefont
  {Ghosez}},\ }\emph {\bibinfo {title} {First-principles study of the
  dielectric and dynamical properties of barium titanate}},\ \href@noop {}
  {\bibinfo {type} {Ph. d. thesis}},\ \bibinfo  {school} {Universiti\'e
  Catholique de Louvain} (\bibinfo {year} {1997}),\ \bibinfo {note} {available
  on-line in http://www.phythema.ulg.ac.be/Books/PhD-Ph.Ghosez.pdf}\BibitemShut
  {NoStop}%
\bibitem [{\citenamefont {Rahman}\ \emph {et~al.}(2019)\citenamefont {Rahman},
  \citenamefont {Haque},\ and\ \citenamefont {Hossain}}]{Rahman_2019}%
  \BibitemOpen
  \bibfield  {author} {\bibinfo {author} {\bibfnamefont {M.~T.}\ \bibnamefont
  {Rahman}}, \bibinfo {author} {\bibfnamefont {E.}~\bibnamefont {Haque}}, \
  and\ \bibinfo {author} {\bibfnamefont {M.~A.}\ \bibnamefont {Hossain}},\
  }\href {\doibase https://doi.org/10.1016/j.jallcom.2018.12.339} {\bibfield
  {journal} {\bibinfo  {journal} {Journal of Alloys and Compounds}\ }\textbf
  {\bibinfo {volume} {783}},\ \bibinfo {pages} {593} (\bibinfo {year}
  {2019})}\BibitemShut {NoStop}%
\bibitem [{\citenamefont {Muchiri}\ \emph {et~al.}(2019)\citenamefont
  {Muchiri}, \citenamefont {Mwalukuku}, \citenamefont {Korir}, \citenamefont
  {Amolo},\ and\ \citenamefont {Makau}}]{MUCHIRI2019489}%
  \BibitemOpen
  \bibfield  {author} {\bibinfo {author} {\bibfnamefont {P.}~\bibnamefont
  {Muchiri}}, \bibinfo {author} {\bibfnamefont {V.}~\bibnamefont {Mwalukuku}},
  \bibinfo {author} {\bibfnamefont {K.}~\bibnamefont {Korir}}, \bibinfo
  {author} {\bibfnamefont {G.}~\bibnamefont {Amolo}}, \ and\ \bibinfo {author}
  {\bibfnamefont {N.}~\bibnamefont {Makau}},\ }\href {\doibase
  https://doi.org/10.1016/j.matchemphys.2019.03.001} {\bibfield  {journal}
  {\bibinfo  {journal} {Materials Chemistry and Physics}\ }\textbf {\bibinfo
  {volume} {229}},\ \bibinfo {pages} {489 } (\bibinfo {year}
  {2019})}\BibitemShut {NoStop}%
\bibitem [{\citenamefont {Chen}\ \emph {et~al.}(2011)\citenamefont {Chen},
  \citenamefont {Niu}, \citenamefont {Li},\ and\ \citenamefont
  {Li}}]{vickers_hardness}%
  \BibitemOpen
  \bibfield  {author} {\bibinfo {author} {\bibfnamefont {X.-Q.}\ \bibnamefont
  {Chen}}, \bibinfo {author} {\bibfnamefont {H.}~\bibnamefont {Niu}}, \bibinfo
  {author} {\bibfnamefont {D.}~\bibnamefont {Li}}, \ and\ \bibinfo {author}
  {\bibfnamefont {Y.}~\bibnamefont {Li}},\ }\href {\doibase
  https://doi.org/10.1016/j.intermet.2011.03.026} {\bibfield  {journal}
  {\bibinfo  {journal} {Intermetallics}\ }\textbf {\bibinfo {volume} {19}},\
  \bibinfo {pages} {1275 } (\bibinfo {year} {2011})}\BibitemShut {NoStop}%
\bibitem [{\citenamefont {Born}(1940)}]{born_1940}%
  \BibitemOpen
  \bibfield  {author} {\bibinfo {author} {\bibfnamefont {M.}~\bibnamefont
  {Born}},\ }\href {\doibase 10.1017/S0305004100017138} {\bibfield  {journal}
  {\bibinfo  {journal} {Mathematical Proceedings of the Cambridge Philosophical
  Society}\ }\textbf {\bibinfo {volume} {36}},\ \bibinfo {pages} {160–172}
  (\bibinfo {year} {1940})}\BibitemShut {NoStop}%
\bibitem [{\citenamefont {Hill}(1952)}]{Hill}%
  \BibitemOpen
  \bibfield  {author} {\bibinfo {author} {\bibfnamefont {R.}~\bibnamefont
  {Hill}},\ }\href {\doibase 10.1088/0370-1298/65/5/307} {\ \textbf {\bibinfo
  {volume} {65}},\ \bibinfo {pages} {349} (\bibinfo {year} {1952})}\BibitemShut
  {NoStop}%
\bibitem [{\citenamefont {Moss}(1985)}]{TMOSS}%
  \BibitemOpen
  \bibfield  {author} {\bibinfo {author} {\bibfnamefont {T.~S.}\ \bibnamefont
  {Moss}},\ }\href {\doibase https://doi.org/10.1002/pssb.2221310202}
  {\bibfield  {journal} {\bibinfo  {journal} {physica status solidi (b)}\
  }\textbf {\bibinfo {volume} {131}},\ \bibinfo {pages} {415} (\bibinfo {year}
  {1985})}\BibitemShut {NoStop}%
\bibitem [{\citenamefont {Garc\'{\i}a-Fern\'andez}\ \emph
  {et~al.}(2016)\citenamefont {Garc\'{\i}a-Fern\'andez}, \citenamefont
  {Wojde\l{}}, \citenamefont {\'I\~niguez},\ and\ \citenamefont
  {Junquera}}]{second_principles}%
  \BibitemOpen
  \bibfield  {author} {\bibinfo {author} {\bibfnamefont {P.}~\bibnamefont
  {Garc\'{\i}a-Fern\'andez}}, \bibinfo {author} {\bibfnamefont {J.~C.}\
  \bibnamefont {Wojde\l{}}}, \bibinfo {author} {\bibfnamefont {J.}~\bibnamefont
  {\'I\~niguez}}, \ and\ \bibinfo {author} {\bibfnamefont {J.}~\bibnamefont
  {Junquera}},\ }\href {\doibase 10.1103/PhysRevB.93.195137} {\bibfield
  {journal} {\bibinfo  {journal} {Phys. Rev. B}\ }\textbf {\bibinfo {volume}
  {93}},\ \bibinfo {pages} {195137} (\bibinfo {year} {2016})}\BibitemShut
  {NoStop}%
\bibitem [{\citenamefont {Sifuna}\ \emph
  {et~al.}(2020{\natexlab{b}})\citenamefont {Sifuna}, \citenamefont
  {Garc\'{\i}a-Fern\'andez}, \citenamefont {Manyali}, \citenamefont {Amolo},\
  and\ \citenamefont {Junquera}}]{mrs-advances-james}%
  \BibitemOpen
  \bibfield  {author} {\bibinfo {author} {\bibfnamefont {J.}~\bibnamefont
  {Sifuna}}, \bibinfo {author} {\bibfnamefont {P.}~\bibnamefont
  {Garc\'{\i}a-Fern\'andez}}, \bibinfo {author} {\bibfnamefont {G.~S.}\
  \bibnamefont {Manyali}}, \bibinfo {author} {\bibfnamefont {G.}~\bibnamefont
  {Amolo}}, \ and\ \bibinfo {author} {\bibfnamefont {J.}~\bibnamefont
  {Junquera}},\ }\href {\doibase 10.1557/adv.2020.111} {\bibfield  {journal}
  {\bibinfo  {journal} {MRS Advances}\ }\textbf {\bibinfo {volume} {5}},\
  \bibinfo {pages} {2281–2290} (\bibinfo {year}
  {2020}{\natexlab{b}})}\BibitemShut {NoStop}%
\end{thebibliography}%
\newpage

\begin{table*}[b]
	\caption[ ]{$\varepsilon_1$ (0), $\varepsilon_2$ ($\omega$),$\alpha$ ($\omega$), R (0) and their maximum values}
	\centering
\begin{tabular}{cccccccc}
	& $\varepsilon_1$ (0)&Max. value of $\varepsilon_1$ ($\omega$)   & Max. value of $\varepsilon_2$ ($\omega$)  &Max. value of $\alpha$($\omega$)  &R(0)\%&Max. value of R ($\omega$)\% \\
		& &(at energy (eV))  &(at energy (eV))  &(at energy (eV)) &&(at energy (eV)) \\
	\hline \\
Ca$_3$BiN	&9.1434 &11.6204 (2.0420) &13.5679 (4.0841)  &$1.876\times10^5$ cm$^{-1}$ (4.885) & 25&50 (4.965)\\
Sr$_3$BiN	&9.9135 &14.0234 (2.2422) &19.5990 (4.4444)  &$2.424\times10^5$ cm$^{-1}$ (4.565) & 27&57 (4.565)\\
Ba$_3$BiN	&11.2508 &13.9910 (1.0010) &13.3950 (3.2032)  &$1.946\times10^5$ cm$^{-1}$ (6.246) & 29&49 (6.807)\\
	\hline
\end{tabular}
\label{table:max_es_at_given_energy}
\end{table*}
\end{document}